\documentclass[twocolumn,superscriptaddress,amsmath,amssymb, aps, prl]{revtex4-2}

\usepackage{float}
\usepackage{subfigure} 
\usepackage{bm}                 
\usepackage{graphicx}% Include figure files
\usepackage{dcolumn}% Align table columns on decimal point
\usepackage{bm}% bold math
\usepackage[colorlinks,citecolor=blue,linkcolor=blue,urlcolor=blue]{hyperref}% add hypertext capabilities

\begin{document}

\title{Electronic-vibrational dynamics and coherence in x-ray transient absorption of N$_2^+$ induced by strong-field ionization}

\author{Jing Zhao} 
\email{jzhao@nudt.edu.cn}
\author{Guangru Bai}%
\author{Qian Zhang}
\author{Bin Zhang}%
\author{Wenkai Tao}%
\author{Qianyu Qiu}%
\author{Hongbin Lei}%
\author{Yue Lang}%
\author{Jinlei Liu}%
\author{Xiaowei Wang}%
\affiliation{Department of Physics, National University of Defense Technology, Changsha 410073, China}
\author{Zengxiu Zhao}
\email{zhaozengxiu@nudt.edu.cn}
\affiliation{Department of Physics, National University of Defense Technology, Changsha 410073, China}
\affiliation{Hunan Key Laboratory of Extreme Matter and Applications, National University of Defense Technology, Changsha 410073, China}

\date{\today}

\begin{abstract}
Attosecond transient absorption spectroscopy (ATAS) is becoming an indispensable and powerful tool in the emerging field of attochemistry, while the interpretation of measurements often requires full considerations of the coupling among various freedoms of motion.  Here we develop the ionization-coupling model to incorporate the transient absorption and  explore the coupled electronic-vibrational  dynamics of nitrogen ions induced by strong-field ionization (SFI), which has been investigated in the recent transient x-ray K-edge absorption experiment [PRL 129, 123002 (2022)].  It is found the coherent vibrational wave packet on the involved electronic state is created with broad distribution of vibrational levels which leads to the spectral overlap on the K-edge absorption. By identifying the contributions of each electronic state, the study provides a different interpretation revealing the significant role of the excited state $A^2\Pi_u$ arising from the electronic-vibrational coupling in strong laser fields. We uncover new features of absorption from forbidden transitions during the laser pulse and confirm the vibronic coherence induced modulations of absorbance after SFI.  A new scheme is proposed to avoid the spectral overlap and determine the population among the states which is crucial to resolve the debate on nitrogen air lasing. The work lays down the framework to research the ionic coherence in ATAS and offers valuable insights into the intricate interplay between electronic and vibrational dynamics.

\end{abstract}

\maketitle

Strong-field ionization (SFI) of molecules creates transient non-stationary ionic states that evolve coherently under the remaining laser field. The SFI-induced electronic, vibrational and rotational excitation, as well as the quantum coherence make the molecular ions unique platform for investigating and manipulating the chemical reactions \cite{worner2023nature,leone2023science,worner2022np,worner2021science,leone2019science,HanM2023L}, and extraordinary radiations such as air lasing and supercontinuum generation \cite{Dogariu11,Yao2011,zhao2022nc}. Determining the intricate multiple-freedom coupled sub-cycle ionic dynamics upon SFI is challenging both theoretically and experimentally \cite{goulielmakis2010real,sabbar2017np,saito2019real,Smirnova09S, Rohringer2009,Pabst2012A2,zhang2020sub,zhao2024nc}, and key questions such as the mechanism of lasing from SFI-created nitrogen molecular ions remain unaddressed since the discovery \cite{Yao2011}. 

Subjecting nitrogen molecules to strong laser fields, the N$_2^+$ can be created in the electronic states $X^2\Sigma^+_g$, $A^2\Pi_u$ and $B^2\Sigma^+_u$.  Laser-like emissions from N$_2^+$ are observed in the forward direction at wavelengths of 391 and 428 nm, corresponding to transitions from the upper excited state $B^2\Sigma^+_u (\nu\!=\!0)$ to the lower ground state $X^2\Sigma^+_g (\nu\!=\!0,1)$. Population inversion between $B^2\Sigma^+_u$ and $X^2\Sigma^+_g$ states is the most straightforward explanation for the lasing effect \cite{yao2016population,xu2015sub}. But how is the population inversion formed? as the tunneling ionization theory predicts less ionization rate into $B^2\Sigma^+_u$ state due to its larger ionization energy. One possible explanation is that the recollision of photoelectron with the cation might lead to population transfer from $X^2\Sigma^+_g$ to $B^2\Sigma^+_u$ states \cite{liu15prl}. However, other experiments show lasing persists even with circularly polarized driving pulses, which is against the recollision picture \cite{corkum18prl}. Later it is realized that the electronic-vibrational coupling is crucial as the single-photon transition from $X^2\Sigma^+_g$ to $A^2\Pi_u$ states can significantly deplete the population on the ionic ground state, leading to possible population inversion between relevant X-B vibrational states \cite{yao2016population,xu2015sub,xu2020prl}. It is further demonstrated that the upper $B^2\Sigma^+_u$ state could be effectively populated through instantaneous polarization of the new born cation in the field \cite{zhang2020sub} and multi-photon resonant transitions facilitated by dynamic Stark-shift \cite{zhao2022nc}.
On the other hand, SFI could create ions in partial coherence \cite{goulielmakis2010real,sabbar2017np,saito2019real}, which might lead to lasing without inversion, as the electronic-vibrational levels of $X^2\Sigma^+_g$, $A^2\Pi_u$ and $B^2\Sigma^+_u$ form V-type configurations \cite{Mysyrowicz2019}. It is also possible that rotational coherence comes into play which causes transient population inversion due to the different rotational constants of $B^2\Sigma^+_u$ and $X^2\Sigma^+_g$ states \cite{arissian2018pra, richter2020rotational, xie2020role,erik2021pra}.  However this interpretation seems to be difficult to explain the lasing efficiency at a pump wavelength of 800 nm, as demonstrated in \cite{Ando2019Rota,yao2016population,xu2015sub,Tikhonchuk2021NJ}. Nevertheless, uncovering the underlying mechanism of lasing in N$^+_2$ relies on state-resolved population and coherence measurements during and after strong laser fields with attosecond temporal resolution.

Attosecond pulses, with short durations and broad bandwidth, offer unique advantages for transient measurements. Particularly, attosecond transient absorption spectroscopy (ATAS) has been applied to probe the laser-induced ionic dynamics initiated by SFI  \cite{goulielmakis2010real,sabbar2017np,zhao2024nc,saito2019real,golubev2021prl,pfeifer2023sa}. In ATAS, the strong laser pulses create transient non-stationary ionic states, and the time-delayed attosecond pulses probe the population and coherence of the transient states through one-photon absorption from the inner shell to the valence shell. 
Especially, ultrafast x-ray absorption spectroscopy (XAS) was employed to inspect N$^+_2$ population distributions \cite{kleine2022electronic}. 
However, in presence of the strong laser field, the SFI induced electronic, vibrational and rotational states are strongly coupled which leads to blurred and extremely complicated absorption spectra. To interpret the observations regarding the N$_2^+$ lasing mechanism, and more importantly, to fully explore the capabilities of transient XAS involving SFI, a comprehensive theory is both indispensable and urgently needed \cite{Calegari2016jpb,Nisoli19atto,leone2022jcp}, which requires full consideration of the coupling among various degrees of freedoms.

\begin{figure}[ht]
  \centering %
  \includegraphics[width=0.4\textwidth]{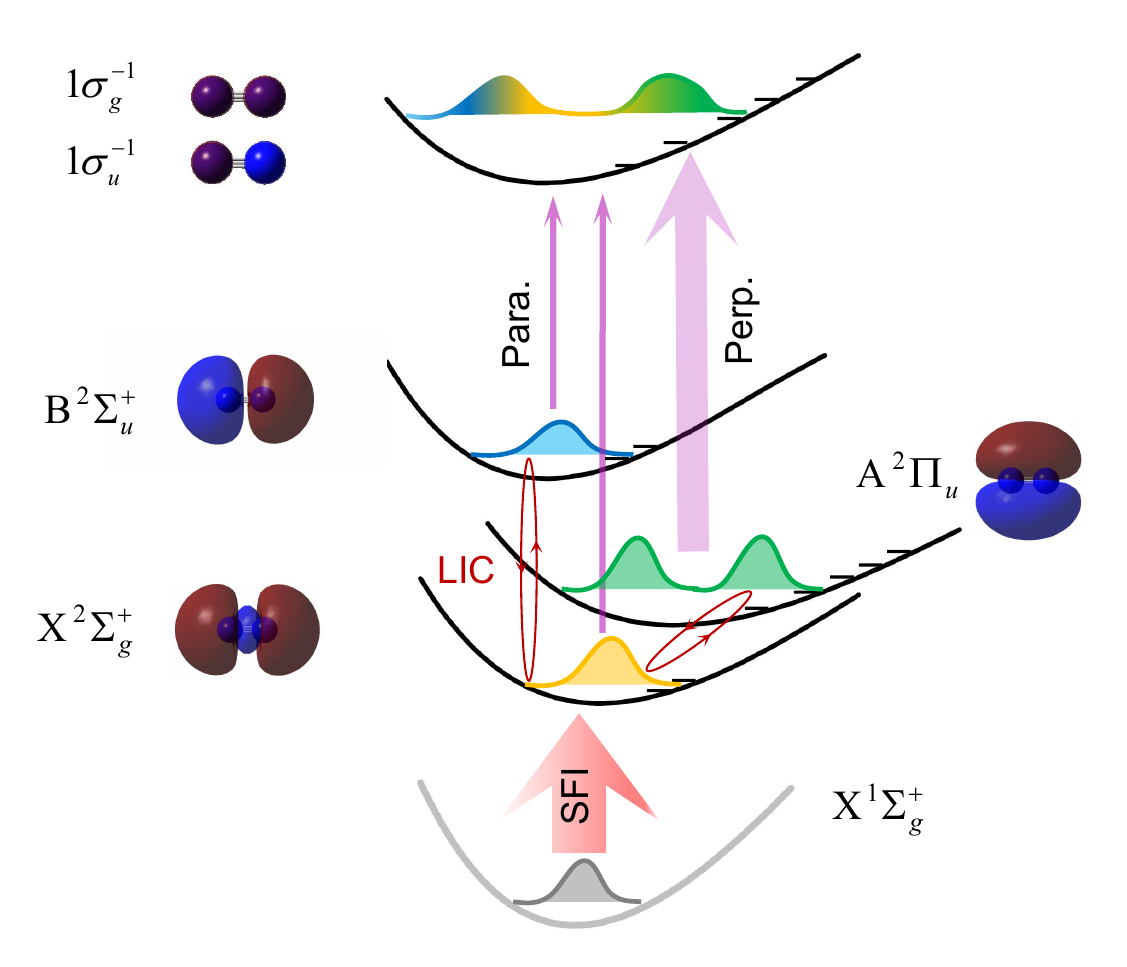}
\caption{Overview of the pump-probe scheme for N$_2$: SFI creates electronic populations in the first three electronic states of N$_2^+$, namely X, A, and B, each with its respective vibrational wavepackets. The laser-induced coupling (LIC, indicated by circles with arrows) leads to the redistribution of electronic population among theses ionic states. The electronic-vibrational dynamics and coherence are probed by XAS through parallel or perpendicular transitions from the nitrogen K-edge.}
\label{pump-probe}
\end{figure}

In this Letter, we develop the ionization-coupling model \cite{zhang2020sub} to incorporate transient absorption and explore the coupled electronic-vibrational dynamics induced by SFI. The relevant processes are illustrated in Fig.~\ref{pump-probe}, where SFI creates ions in the electronic states $X^2\Sigma^+_g$, $A^2\Pi_u$ and $B^2\Sigma^+_u$, each with its respective vibrational wave packet. The laser-induced electronic-vibrational coupling redistributes the population among the electronic and vibrational states in N$_2^+$. The evolution of population and coherence of participated states are probed by XAS via transitions from the nitrogen K-edge. Our calculations reproduce well the experimental time-resolved absorption spectra of N$^+_2$ \cite{kleine2022electronic}. It is found that the laser-induced electronic-vibrational coupling leads to the coherent vibrational wave packet with broad distribution of vibrational levels, which results in spectral overlap with the K-edge absorption. By identifying the contributions of each electronic state, we verify a remarkably high population of the $A^2\Pi_u$ state. Our study emphasizes the importance of electronic-vibrational coupling in strong laser field \cite{golubev2021prl}. Moreover, we have observed both forbidden transitions during the laser pulse and absorbance modulation after the laser pulsed, which are induced by vibronic coherence. A new scheme is proposed to avoid the spectral overlap and determine the population among the states, which is crucial for resolving the debate on nitrogen air lasing.

The evolution of the ionic density matrix under strong laser field is given by 
\begin{equation}
  \dfrac{d\rho_{i\nu}^+}{dt}=-\dfrac{i}{\hbar}\left[H(t),\rho_{i\nu}^+\right]+\\
  \left(\dfrac{d\rho_{i\nu}^+}{dt}\right)_{\rm ion}+\left(\dfrac{d\rho_{i\nu}^+}{dt}\right)_{\rm decay},
  \label{eq:one}
\end{equation}
where $\rho^{+}_{i\nu}$ is the ionic density matrix element, with $i$ and $\nu$ denoting the electronic states and vibrational levels. At every instant, SFI injects coherent vibrational wave packets into the electronic states $X^2\Sigma^+_g$, $A^2\Pi_u$ and $B^2\Sigma^+_u$. The injection rate is calculated using the MO-ADK theory \cite{tong2002theory} with the SLIMP package \cite{zhang2015slimp}. The vibrational wave packets injected into different electronic states are assumed to be incoherent since the SFI destroys the electronic coherence \cite{goulielmakis2010real}. The subsequent laser-ion interaction establish coherence among different vibrational wave packets by laser-induced coupling. The ionic hamiltonian $H(t)$ includes the interaction between the ion and the pump-probe field, described with the electric dipole approximation. The tensor transition dipole moment is constructed by weighting the electronic transition dipole moment $a_{i,j}$ with their respective Franck-Condon factors $\mu_{i\nu,j\nu'}=\langle\varphi_{i,\nu}|a_{i,j}|\varphi_{j,\nu'}\rangle$, describing the field-induced coupling among different states (for details, see SM Section 1.1). By solving Eq.~(\ref{eq:one}) numerically, we obtain the time-dependent dipole moment at the pump-probe time delay $t_d$,
\begin{equation}
\bm{d}(\theta,t,t_d)=\sum_{i,j}\sum_{\nu,\nu'}\rho^+_{i\nu,j\nu'}(\theta,t,t_d)\mu_{i\nu,j\nu'},
\end{equation}
at the alignment angle $\theta$ between the molecular axis and the polarization direction of the laser field. The density matrix element $\rho^+_{i\nu,j\nu'}(\theta,t,t_d)$ represents the coherence between different states. The transient absorption spectra are calculated as in \cite{wirth2011synthesized,santra2011pra},
\begin{equation}
  \sigma(\theta,\omega,t_d)=\dfrac{4\pi\alpha\omega}{|{\bm \varepsilon}(\omega)|^2}{\rm Im}\left[{\bm d}(\theta,\omega,t_d)\cdot{\bm \varepsilon}^*(\omega)\right],
  \label{eq:two}
\end{equation}
where ${\bm \varepsilon}(\omega)$ and ${\bm d}(\theta,\omega,t_d)$ represent the Fourier transforms of the electric field and the time-dependent dipole moment, respectively.
In our calculations, we use an 800 nm, $4.5\times10^{14}$\,W/cm$^2$ infrared laser pulse with duration of 50\,fs and the x-ray pulse centered at 393 eV, with $10^{11}$\,W/cm$^2$ intensity and and 200 as duration. Both pump and probe pulses are linearly polarized and parallel. The absorption spectra are independent of the x-ray pulse duration as long as the spectra remain continuous (for details, see Section 2.2 in SM).
We calculate the absorption spectra for various molecular alignments and integrate over all angles to account for alignment dynamics (for details, see SM Section 3). 

\begin{figure}[ht]
  \centering %
  \includegraphics[width=0.45\textwidth]{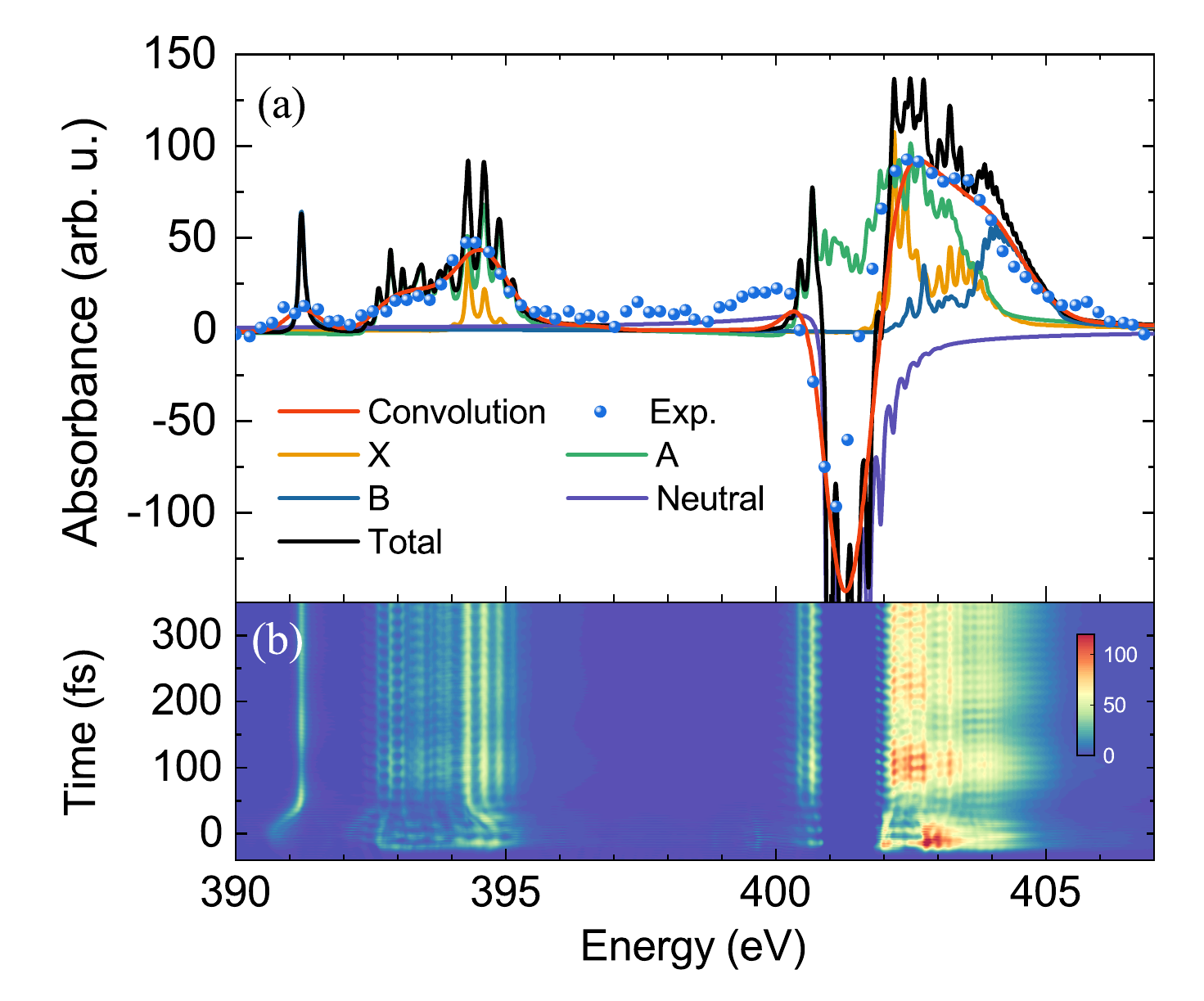}
\caption{(a) The calculated time-integrated XAS of N$_2^+$ (red line) is compared with the experimental results from Ref. \cite{kleine2022electronic} (blue dots)). Contributions from the three electronic state are also shown separately: $X^2\Sigma^+_g$ (orange line), $A^2\Pi_u$ (green line) and $B^2\Sigma^+_u$ (blue line). (b) The transient XAS of nitrogen molecules exposed to strong laser pulses.}
\label{figXAS}
\end{figure}

The calculated transient K-edge absorption spectra of N$_2^+$ are presented in Fig.~\ref{figXAS}(b) as a function of the pump-probe time delay, with positive delays indicating that the laser pulse precedes the x-ray pulse. After the time delay of 75\,fs, N$_2^+$ evolves freely.
The slow oscillations in the transient absorption spectra with a period of hundreds of femtoseconds, originate from the laser-induced molecular alignment effects.
By integrating the absorption spectra over a 140\,fs time interval within the free evolution period, we present the time-integrated absorption spectra in Fig.~\ref{figXAS}(a). The narrow peak around 391.2\,eV results from the resonant transition from $1\sigma_g$ state to $B^2\Sigma^+_u$ state. The transition from $1\sigma_g$ state to $A^2\Pi_u$ state contributes to the absorption spectra in the energy range from 392.5\,eV to 395.5\,eV, spectrally overlapping with the transition from $1\sigma_u$ state to $X^2\Sigma^+_g$ state. Specifically, the peak at 394.4eV, attributed to the transition $1\sigma_u\rightarrow\,X^2\Sigma^+_g$ as described in \cite{kleine2022electronic}, is also significantly contributed by the vibrational-progressed transitions of $1\sigma_g\rightarrow\,A^2\Pi_u$. Additionally, for the satellite transitions from $1\sigma_u$ to the lowest unoccupied $1\pi_g$ with vacancies remaining in the $X^2\Sigma^+_g$ or $A^2\Pi_u$ states, spectral overlap is observed in the energy range of 402-404\,eV.  

The discrete peaks observed in the calculated XAS spectra suggest contributions from multiple vibrational levels, which are not resolved in the experiments due to the limited spectral resolution. To simulate the spectral broadening, we convolute the total absorption spectra with the energy width of 0.3\,eV (red curve). The calculations reproduce well the measurements (blue dots) across the entire absorption spectra of N$_2^{+}$. This agreement between our calculations and measurements validates the precision of our theoretical approach and highlights its potential for exploring the intricate interplay between electronic and vibrational dynamics.

\begin{figure}[ht]
  \centering %
  \includegraphics[width=0.45\textwidth]{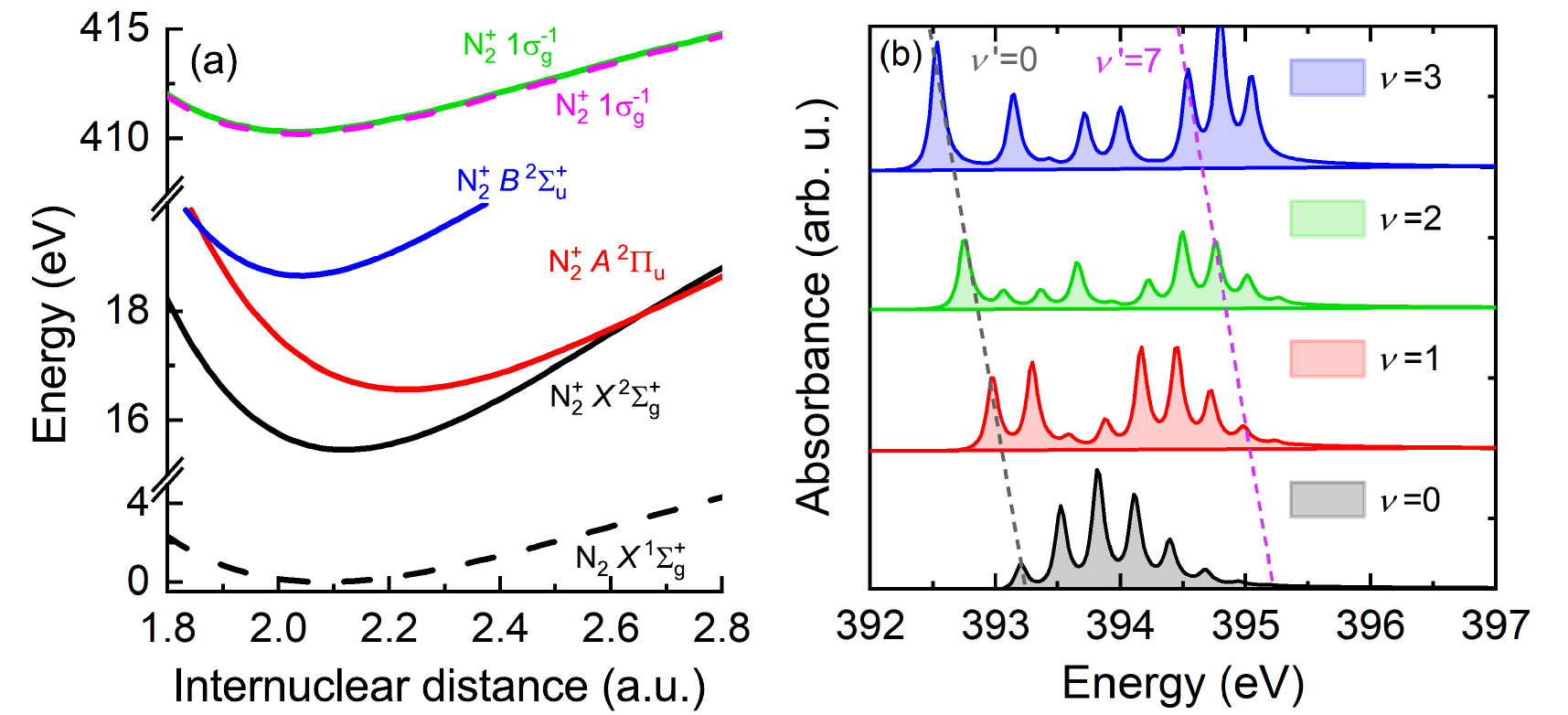}
\caption{(a) The relevant potential energy curves for N$_2$ and N$_2^{+}$. (b) Time-integrated XAS with transitions between $A^2\Pi_u$ state and $1\sigma_g^{-1}$ core-hole state with vibrational levels $\nu=0\sim3$. The dashed lines indicate transition energies to $1\sigma_g^{-1}$ core-hole state with vibrational levels $\nu'=0$ and $\nu'=7$.}
   \label{PEC}
\end{figure}

To explain the spectral extension in the K-edge absorption from $A^2\Pi_u$ state, we present the relevant potential energy curves (PECs) in Fig.~\ref{PEC}(a). Calculation details and comparison with previous results are provided in Section 1.2 of the SM \cite{werme1973fine, ehara2006symmetry, lindblad2020x}. 
In the presence of strong laser fields, $A^2\Pi_u$ state can be efficiently populated through electronic-vibrational coupling with one-photon resonant transition from the ionic ground state $X^2\Sigma^+_g$ \cite{zhang2020sub,zhao2022nc}. Interestingly, the significant variation in the equilibrium nuclear distance between the $X^2\Sigma^+_g$ and $A^2\Pi_u$ states causes a broad distribution of vibrational levels in the latter. Additionally, the electronic-vibrational coupling redistributes the population toward higher vibrational levels, leading to a deviation from the Franck-Condon distribution (see in Section 2.4 of the SM).
On the contrary, for the K-edge absorption from  $A^2\Pi_u$ state to  $1\sigma_g^{-1}$ core-hole state, the change of nuclear separation is even larger, and thus much more vibrational levels of the K-shell are involved. The combination of these effects results in the final spectral extension in the K-edge absorption spectra from $A^2\Pi_u$ state, which is overlooked in \cite{kleine2022electronic}. The K-edge absorption spectra from $A^2\Pi_u$ state, shown in Fig.~\ref{PEC}(b), exhibit an absorption energy range spanning from 392.5 eV to 395.5 eV.
Notably, for the vibrational level $\nu=3$, x-ray absorption populates over ten vibrational levels of the $1\sigma_g^{-1}$ core-hole state, resulting in absorption energy spread of nearly 3\,eV. 
This confirms the crucial role of laser-induced electronic-vibrational coupling in exploring the dynamics of the K-edge absorption spectra \cite{worner2021science,loh2017nc}, which is strongly relies on the accuracy of the PECs. 
Here, we verify that the experimental measurements indeed include significant contributions from $A^2\Pi_u$ state, which was previously underestimated in the theoretical analysis \cite{kleine2022electronic}.

\begin{figure}[ht]
  \centering %
   \includegraphics[width=0.45\textwidth]{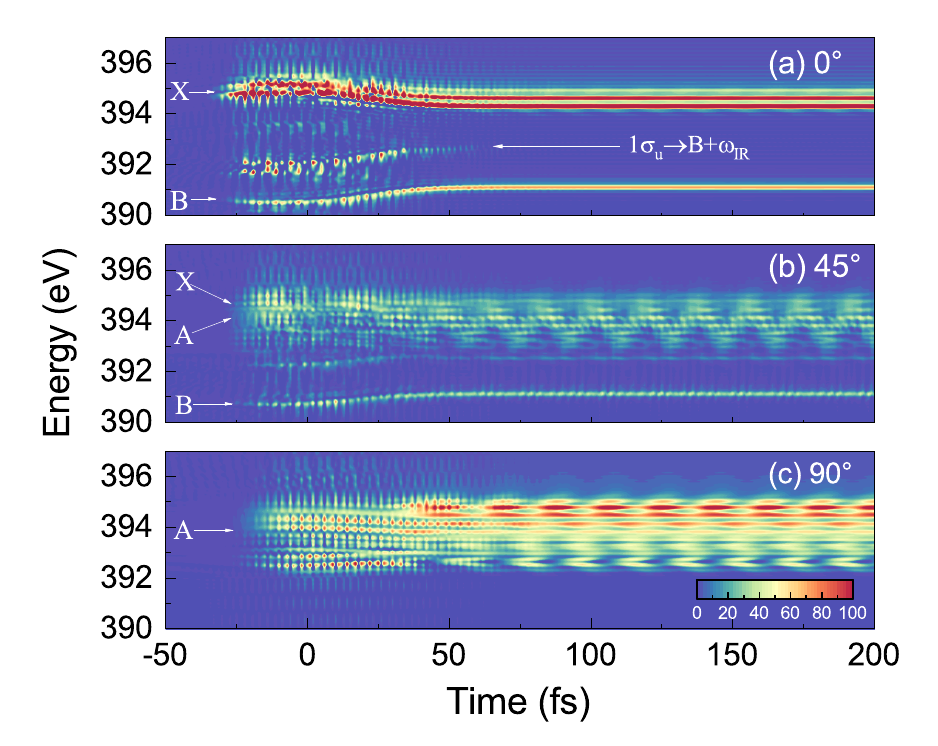}
   \caption{Transient x-ray absorption spectra are presented for different alignment angles between the molecular axis and the polarization direction of the pump-probe field: (a) $0^{\circ}$, (b) $45^{\circ}$, and (c) $90^{\circ}$.}
  \label{XAS_angle}
\end{figure}

Notably, the population dynamics and vibronic coherence can be explored in the K-edge transient absorption spectra, as shown in Fig.~\ref{XAS_angle}. The results are sensitive to the alignment angles between the molecular axis and the polarization direction of the pump-probe field. When the pulse is polarized parallel to the molecular axis, the absorption spectra are primarily dominated by the absorption from $X^2\Sigma^+_g$ and $B^2\Sigma^+_u$ states, while absorption from $A^2\Pi_u$ state is forbidden due to symmetry. The absorption from the $B^2\Sigma^+_u$ state includes the vibrational transitions of 0-0 and 1-1 with the same transition energy of 391.2\,eV. Absorption from $X^2\Sigma^+_g$ state relates to the transitions from the vibrational level $\nu=0$ to $\nu'=0\sim2$.
For the perpendicular alignment, the absorption spectra are mainly contributed by the absorption from $A^2\Pi_u$ state, 
and the absorption from $X^2\Sigma^+_g$ state is not allowed with the perpendicularly polarized x-ray pulse. With the alignment angle of $45^{\circ}$, absorptions from all three electronic states are allowed. 

Within the strong laser pulse, the absorption energies from $X^2\Sigma^+_g$ and $B^2\Sigma^+_u$ states exhibit significant upward and downward shifts, respectively, due to the AC Stark effect arising from laser-induced coupling \cite{zhao2022nc,bakos1977ac}. The rapid oscillations with half-period of the laser pulse originate from the transient polarization of the electronic states under the strong laser field, as observed in the attosecond XAS of atoms \cite{goulielmakis2010real,sabbar2017np,zhao2024nc}. 
Moreover, it is intriguing to note that absorptions from forbidden transitions during the laser pulse \cite{chini2013sr} also manifest in the XAS of N$_2^+$. For instance, the $B^2\Sigma^+_u$ state is not accessible from the $1\sigma_u$ state through single-photon absorption. However, in presence of the laser pulse, this transition can occur by absorbing an additional laser photon, {\rm i.e.}, to the laser-dressed state 1$\sigma_u\rightarrow$B$+\omega_{IR}$, as indicated in Fig.~\ref{XAS_angle}(a). These new features resemble Floquet engineering, which aims to dynamically tailor the electronic properties of solid-state materials \cite{zhou23nature}, underscoring the intricate interplay between the laser field and the x-ray pulse interacting with deep-lying core-shell states.

\begin{figure}[ht]
  \centering %
  \includegraphics[width=0.45\textwidth]{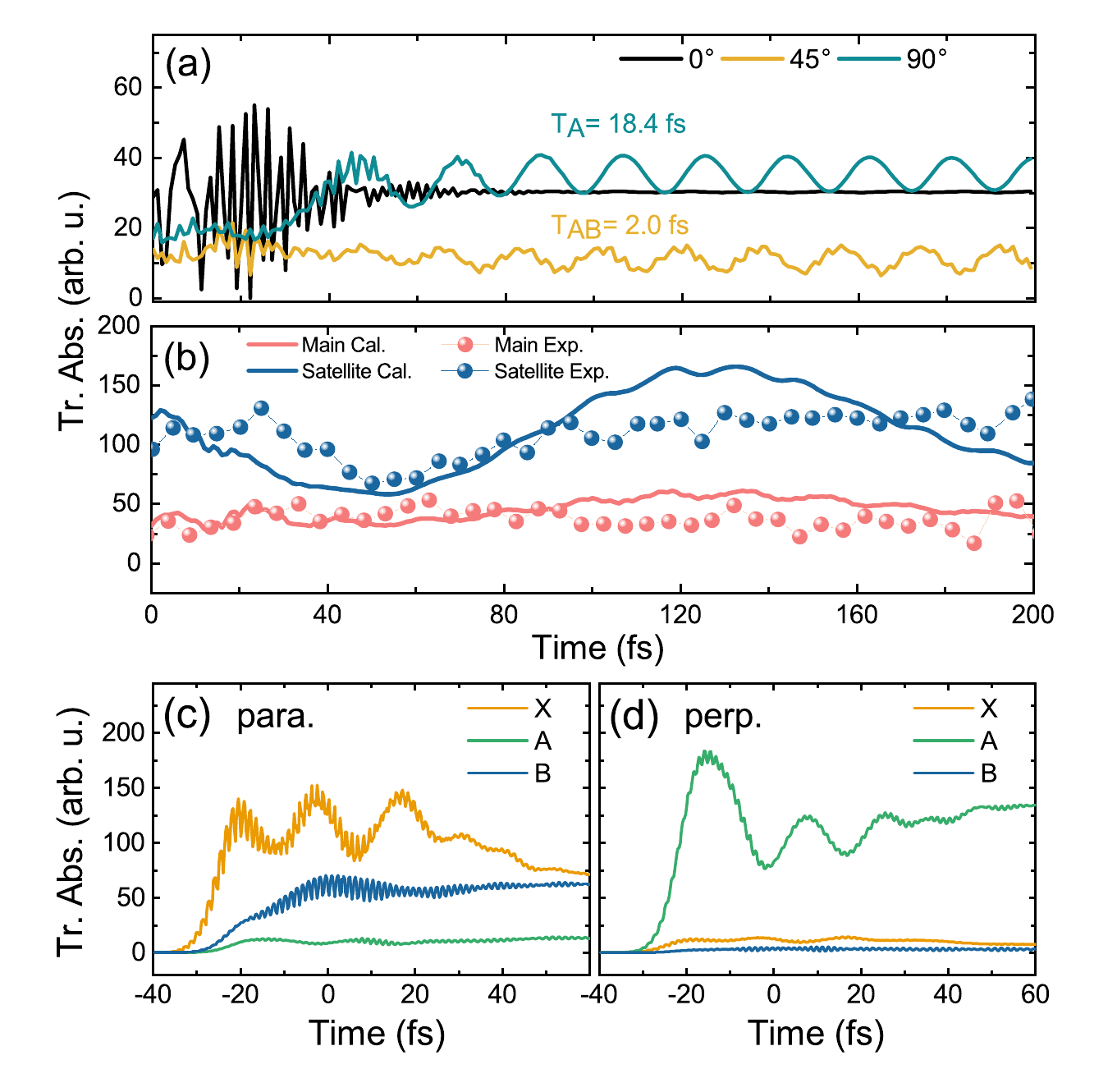}
\caption{(a) The temporal evolutions of XAS integrated from 393.7 eV to 396 eV at alignment angles of $0^{\circ}$, $45^{\circ}$, and $90^{\circ}$. (b) The spectrally integrated absorbance for mainline (red) and satellite line (blue) transitions is compared with experimental results (points) from \cite{kleine2022electronic}. 
(c) and (d) depict electronic population dynamics for molecules pre-aligned at $45^{\circ}$, probed with x-ray pulses parallel and perpendicular to the molecular axis, respectively.}
\label{Abs_t}
\end{figure}

Quantum beats in the K-edge absorbance indicate the existence of vibronic coherence.  
Fig.~\ref{Abs_t}(a) presents the integrated absorption spectra from 393.7 eV to 396 eV, including transitions from both $X^2\Sigma^+_g$ and $A^2\Pi_u$ states. At the alignment angle of $90^{\circ}$, the 18.4\,fs oscillation period is attributed to the coherent vibrational wave packet of $A^2\Pi_u$ state. At $45^{\circ}$, additional rapid oscillations emerge due to interference between $A^2\Pi_u$ and $B^2\Sigma^+_u$ states, both transitioning to the $1\sigma_g^{-1}$ core-hole state. The rotational coherence is evident with a minimum at around 50\,fs when integrating the transient XAS spectra from 402 to 404\,eV. Considering the dynamical alignment effects, our calculations reproduce this minimum within the satellite transitions. 
These the transient absorption oscillations reveal the vibronic and rotational coherence induced by the laser-induced coupling after SFI, which is crucial for charge migrations \cite{worner2022np} and chemical reactions \cite{loh2017nc,vrakking2014np,james2024nature}.

%To compare with experimental measurements, we calculate the total transient XAS considering the time-varying degree of molecular alignment, as shown in Fig.~\ref{Abs_t}(b). Within the transition energy range of 393.7-396 eV, the parallel transition from $X^2\Sigma_g^+$ state and perpendicular transition from $A^2\Pi_u$ state result in the temporal evolution of the absorbance that is insensitive to molecular alignment effects. In contrast, for the satellite transitions in the energy range of 402-404\,eV, the absorbance oscillates with the first minimum around 50\,fs, resembling the measurements. These results demonstrate the existence of the rotational coherence in the transient XAS, which can be probed by choosing appropriate transition energies according to theoretical predictions.

Finally, we propose a scheme to better resolve the population dynamics from XAS by taking advantage of the alignment-dependent transition dipole moments \cite{neumark2020nc}. The population dynamics can be probed separately by varying the polarized direction of the x-ray pulse relative to the molecular axis. Assuming pre-aligned molecules at $45^{\circ}$ with the degree of $\langle\cos^2(\theta)\rangle=0.8$, all three electronic states are populated.
As demonstrated in Fig.~\ref{Abs_t}(c) and (d), the $X^2\Sigma^+_g$ state (orange) and $B^2\Sigma^+_u$ state (blue) can be probed with x-rays polarized parallel to the molecular axis, while the  $A^2\Pi_u$ (green) state is probed by varying the polarization direction perpendicular to the molecular axis. Combining the parallel and perpendicular measurements enables clear resolution transfer from the ionic ground state $X^2\Sigma^+_g$ to the excited state $A^2\Pi_u$ (see Section 3 in SM for detailed calculations). 

In conclusion, we have investigated theoretically the coupled electronic-vibrational dynamics and coherence of nitrogen ions induced by SFI in corroborate with the recent transient x-ray K-edge absorption experiment \cite{kleine2022electronic}. We establish a theoretical approach to simulate the transient absorption and explore the coupled electronic-vibrational dynamics upon SFI, which accurately reproduce the experimental results. By identifying the contributions of each electronic state, it is found that the excited state $A^2\Pi_u$ can be efficiently populated due to the electronic-vibrational coupling in strong laser fields. It indicates that the electronic population inversion occurs at least for certain alignments of nitrogen molecules. The theory helps uncover new features of absorption from forbidden transitions during the laser pulse and confirms that vibronic coherence induces modulations of absorbance after SFI. A new scheme is proposed to determine the population transfer at different probing geometries to avoid the spectral overlap. This work establishes the foundation for studying the photoionization-initiated dynamics and coherence in attochemistry.

\begin{acknowledgments}
This work is supported by the National Key Research and Development Program of China (No. 2019YFA0307703), the National Natural Science Foundation of China (Nos. 12234020, 11874066, 12274461, 11974426), and the Science and Technology Innovation Program of Hunan Province (No. 2022RC1193). The authors appreciate the fruitful discussions with Dr. Erik T.J. Nibbering.

J. Z., G. B., and Q. Z. contributed equally to this work.

%\dots.
\end{acknowledgments}

%\bibliography{apssamp}% Produces the bibliography via BibTeX.

%apsrev4-2.bst 2019-01-14 (MD) hand-edited version of apsrev4-1.bst
%Control: key (0)
%Control: author (8) initials jnrlst
%Control: editor formatted (1) identically to author
%Control: production of article title (0) allowed
%Control: page (0) single
%Control: year (1) truncated
%Control: production of eprint (0) enabled
\providecommand{\noopsort}[1]{}\providecommand{\singleletter}[1]{#1}%
\begin{thebibliography}{49}%
\makeatletter
\providecommand \@ifxundefined [1]{%
 \@ifx{#1\undefined}
}%
\providecommand \@ifnum [1]{%
 \ifnum #1\expandafter \@firstoftwo
 \else \expandafter \@secondoftwo
 \fi
}%
\providecommand \@ifx [1]{%
 \ifx #1\expandafter \@firstoftwo
 \else \expandafter \@secondoftwo
 \fi
}%
\providecommand \natexlab [1]{#1}%
\providecommand \enquote  [1]{``#1''}%
\providecommand \bibnamefont  [1]{#1}%
\providecommand \bibfnamefont [1]{#1}%
\providecommand \citenamefont [1]{#1}%
\providecommand \href@noop [0]{\@secondoftwo}%
\providecommand \href [0]{\begingroup \@sanitize@url \@href}%
\providecommand \@href[1]{\@@startlink{#1}\@@href}%
\providecommand \@@href[1]{\endgroup#1\@@endlink}%
\providecommand \@sanitize@url [0]{\catcode `\\12\catcode `\$12\catcode
  `\&12\catcode `\#12\catcode `\^12\catcode `\_12\catcode `\%12\relax}%
\providecommand \@@startlink[1]{}%
\providecommand \@@endlink[0]{}%
\providecommand \url  [0]{\begingroup\@sanitize@url \@url }%
\providecommand \@url [1]{\endgroup\@href {#1}{\urlprefix }}%
\providecommand \urlprefix  [0]{URL }%
\providecommand \Eprint [0]{\href }%
\providecommand \doibase [0]{https://doi.org/}%
\providecommand \selectlanguage [0]{\@gobble}%
\providecommand \bibinfo  [0]{\@secondoftwo}%
\providecommand \bibfield  [0]{\@secondoftwo}%
\providecommand \translation [1]{[#1]}%
\providecommand \BibitemOpen [0]{}%
\providecommand \bibitemStop [0]{}%
\providecommand \bibitemNoStop [0]{.\EOS\space}%
\providecommand \EOS [0]{\spacefactor3000\relax}%
\providecommand \BibitemShut  [1]{\csname bibitem#1\endcsname}%
\let\auto@bib@innerbib\@empty
%</preamble>
\bibitem [{\citenamefont {Yin}\ \emph {et~al.}(2023)\citenamefont {Yin},
  \citenamefont {Chang}, \citenamefont {Bal{\v c}i{\=u}nas}, \citenamefont
  {Shakya}, \citenamefont {Djorovi{\'c}}, \citenamefont {Gaulier},
  \citenamefont {Fazio}, \citenamefont {Santra}, \citenamefont {Inhester},
  \citenamefont {Wolf},\ and\ \citenamefont {W{\"o}rner}}]{worner2023nature}%
  \BibitemOpen
  \bibfield  {author} {\bibinfo {author} {\bibfnamefont {Z.}~\bibnamefont
  {Yin}}, \bibinfo {author} {\bibfnamefont {Y.-P.}\ \bibnamefont {Chang}},
  \bibinfo {author} {\bibfnamefont {T.}~\bibnamefont {Bal{\v c}i{\=u}nas}},
  \bibinfo {author} {\bibfnamefont {Y.}~\bibnamefont {Shakya}}, \bibinfo
  {author} {\bibfnamefont {A.}~\bibnamefont {Djorovi{\'c}}}, \bibinfo {author}
  {\bibfnamefont {G.}~\bibnamefont {Gaulier}}, \bibinfo {author} {\bibfnamefont
  {G.}~\bibnamefont {Fazio}}, \bibinfo {author} {\bibfnamefont
  {R.}~\bibnamefont {Santra}}, \bibinfo {author} {\bibfnamefont
  {L.}~\bibnamefont {Inhester}}, \bibinfo {author} {\bibfnamefont {J.-P.}\
  \bibnamefont {Wolf}},\ and\ \bibinfo {author} {\bibfnamefont {H.~J.}\
  \bibnamefont {W{\"o}rner}},\ }\bibfield  {title} {\bibinfo {title}
  {Femtosecond proton transfer in urea solutions probed by x-ray
  spectroscopy},\ }\href {https://doi.org/10.1038/s41586-023-06182-6}
  {\bibfield  {journal} {\bibinfo  {journal} {Nature}\ }\textbf {\bibinfo
  {volume} {619}},\ \bibinfo {pages} {749} (\bibinfo {year}
  {2023})}\BibitemShut {NoStop}%
\bibitem [{\citenamefont {Ridente}\ \emph {et~al.}(2023)\citenamefont
  {Ridente}, \citenamefont {Hait}, \citenamefont {Haugen}, \citenamefont
  {Ross}, \citenamefont {Neumark}, \citenamefont {Head-Gordon},\ and\
  \citenamefont {Leone}}]{leone2023science}%
  \BibitemOpen
  \bibfield  {author} {\bibinfo {author} {\bibfnamefont {E.}~\bibnamefont
  {Ridente}}, \bibinfo {author} {\bibfnamefont {D.}~\bibnamefont {Hait}},
  \bibinfo {author} {\bibfnamefont {E.~A.}\ \bibnamefont {Haugen}}, \bibinfo
  {author} {\bibfnamefont {A.~D.}\ \bibnamefont {Ross}}, \bibinfo {author}
  {\bibfnamefont {D.~M.}\ \bibnamefont {Neumark}}, \bibinfo {author}
  {\bibfnamefont {M.}~\bibnamefont {Head-Gordon}},\ and\ \bibinfo {author}
  {\bibfnamefont {S.~R.}\ \bibnamefont {Leone}},\ }\bibfield  {title} {\bibinfo
  {title} {Femtosecond symmetry breaking and coherent relaxation of methane
  cations via x-ray spectroscopy},\ }\href
  {https://doi.org/10.1126/science.adg4421} {\bibfield  {journal} {\bibinfo
  {journal} {Science}\ }\textbf {\bibinfo {volume} {380}},\ \bibinfo {pages}
  {713} (\bibinfo {year} {2023})}\BibitemShut {NoStop}%
\bibitem [{\citenamefont {Matselyukh}\ \emph {et~al.}(2022)\citenamefont
  {Matselyukh}, \citenamefont {Despr\'{e}}, \citenamefont {Golubev},
  \citenamefont {Kuleff},\ and\ \citenamefont {W\"{o}rner}}]{worner2022np}%
  \BibitemOpen
  \bibfield  {author} {\bibinfo {author} {\bibfnamefont {D.~T.}\ \bibnamefont
  {Matselyukh}}, \bibinfo {author} {\bibfnamefont {V.}~\bibnamefont
  {Despr\'{e}}}, \bibinfo {author} {\bibfnamefont {N.~V.}\ \bibnamefont
  {Golubev}}, \bibinfo {author} {\bibfnamefont {A.~I.}\ \bibnamefont
  {Kuleff}},\ and\ \bibinfo {author} {\bibfnamefont {H.~J.}\ \bibnamefont
  {W\"{o}rner}},\ }\bibfield  {title} {\bibinfo {title} {Decoherence and
  revival in attosecond charge migration driven by non-adiabatic dynamics},\
  }\href {https://doi.org/10.1038/s41567-022-01690-0} {\bibfield  {journal}
  {\bibinfo  {journal} {Nature Physics}\ }\textbf {\bibinfo {volume} {18}},\
  \bibinfo {pages} {1206} (\bibinfo {year} {2022})}\BibitemShut {NoStop}%
\bibitem [{\citenamefont {Zinchenko}\ \emph {et~al.}(2021)\citenamefont
  {Zinchenko}, \citenamefont {Ardana-Lamas}, \citenamefont {Seidu},
  \citenamefont {Neville}, \citenamefont {van~der Veen}, \citenamefont
  {Lanfaloni}, \citenamefont {Schuurman},\ and\ \citenamefont
  {W{\"o}rner}}]{worner2021science}%
  \BibitemOpen
  \bibfield  {author} {\bibinfo {author} {\bibfnamefont {K.~S.}\ \bibnamefont
  {Zinchenko}}, \bibinfo {author} {\bibfnamefont {F.}~\bibnamefont
  {Ardana-Lamas}}, \bibinfo {author} {\bibfnamefont {I.}~\bibnamefont {Seidu}},
  \bibinfo {author} {\bibfnamefont {S.~P.}\ \bibnamefont {Neville}}, \bibinfo
  {author} {\bibfnamefont {J.}~\bibnamefont {van~der Veen}}, \bibinfo {author}
  {\bibfnamefont {V.~U.}\ \bibnamefont {Lanfaloni}}, \bibinfo {author}
  {\bibfnamefont {M.~S.}\ \bibnamefont {Schuurman}},\ and\ \bibinfo {author}
  {\bibfnamefont {H.~J.}\ \bibnamefont {W{\"o}rner}},\ }\bibfield  {title}
  {\bibinfo {title} {Sub-7-femtosecond conical-intersection dynamics probed at
  the carbon {K}-edge},\ }\href {https://doi.org/doi:10.1126/science.abf1656}
  {\bibfield  {journal} {\bibinfo  {journal} {Science}\ }\textbf {\bibinfo
  {volume} {371}},\ \bibinfo {pages} {489} (\bibinfo {year}
  {2021})}\BibitemShut {NoStop}%
\bibitem [{\citenamefont {Kobayashi}\ \emph {et~al.}(2019)\citenamefont
  {Kobayashi}, \citenamefont {Chang}, \citenamefont {Zeng}, \citenamefont
  {Neumark},\ and\ \citenamefont {Leone}}]{leone2019science}%
  \BibitemOpen
  \bibfield  {author} {\bibinfo {author} {\bibfnamefont {Y.}~\bibnamefont
  {Kobayashi}}, \bibinfo {author} {\bibfnamefont {K.~F.}\ \bibnamefont
  {Chang}}, \bibinfo {author} {\bibfnamefont {T.}~\bibnamefont {Zeng}},
  \bibinfo {author} {\bibfnamefont {D.~M.}\ \bibnamefont {Neumark}},\ and\
  \bibinfo {author} {\bibfnamefont {S.~R.}\ \bibnamefont {Leone}},\ }\bibfield
  {title} {\bibinfo {title} {Direct mapping of curve-crossing dynamics in {IBr}
  by attosecond transient absorption spectroscopy},\ }\href
  {https://doi.org/10.1126/science.aax0076} {\bibfield  {journal} {\bibinfo
  {journal} {Science}\ }\textbf {\bibinfo {volume} {365}},\ \bibinfo {pages}
  {79} (\bibinfo {year} {2019})}\BibitemShut {NoStop}%
\bibitem [{\citenamefont {Han}\ \emph {et~al.}(2023)\citenamefont {Han},
  \citenamefont {Fedyk}, \citenamefont {Ji}, \citenamefont {Despr\'e},
  \citenamefont {Kuleff},\ and\ \citenamefont {W\"orner}}]{HanM2023L}%
  \BibitemOpen
  \bibfield  {author} {\bibinfo {author} {\bibfnamefont {M.}~\bibnamefont
  {Han}}, \bibinfo {author} {\bibfnamefont {J.}~\bibnamefont {Fedyk}}, \bibinfo
  {author} {\bibfnamefont {J.-B.}\ \bibnamefont {Ji}}, \bibinfo {author}
  {\bibfnamefont {V.}~\bibnamefont {Despr\'e}}, \bibinfo {author}
  {\bibfnamefont {A.~I.}\ \bibnamefont {Kuleff}},\ and\ \bibinfo {author}
  {\bibfnamefont {H.~J.}\ \bibnamefont {W\"orner}},\ }\bibfield  {title}
  {\bibinfo {title} {Observation of nuclear wave-packet interference in
  ultrafast interatomic energy transfer},\ }\href
  {https://doi.org/10.1103/PhysRevLett.130.253202} {\bibfield  {journal}
  {\bibinfo  {journal} {Phys. Rev. Lett.}\ }\textbf {\bibinfo {volume} {130}},\
  \bibinfo {pages} {253202} (\bibinfo {year} {2023})}\BibitemShut {NoStop}%
\bibitem [{\citenamefont {Dogariu}\ \emph {et~al.}(2011)\citenamefont
  {Dogariu}, \citenamefont {Michael}, \citenamefont {Scully},\ and\
  \citenamefont {Miles}}]{Dogariu11}%
  \BibitemOpen
  \bibfield  {author} {\bibinfo {author} {\bibfnamefont {A.}~\bibnamefont
  {Dogariu}}, \bibinfo {author} {\bibfnamefont {J.~B.}\ \bibnamefont
  {Michael}}, \bibinfo {author} {\bibfnamefont {M.~O.}\ \bibnamefont
  {Scully}},\ and\ \bibinfo {author} {\bibfnamefont {R.~B.}\ \bibnamefont
  {Miles}},\ }\bibfield  {title} {\bibinfo {title} {High-gain backward lasing
  in air},\ }\href {https://doi.org/10.1126/science.1199492} {\bibfield
  {journal} {\bibinfo  {journal} {Science}\ }\textbf {\bibinfo {volume}
  {331}},\ \bibinfo {pages} {442} (\bibinfo {year} {2011})}\BibitemShut
  {NoStop}%
\bibitem [{\citenamefont {Yao}\ \emph {et~al.}(2011)\citenamefont {Yao},
  \citenamefont {Zeng}, \citenamefont {Xu}, \citenamefont {Li}, \citenamefont
  {Chu}, \citenamefont {Ni}, \citenamefont {Zhang}, \citenamefont {Chin},
  \citenamefont {Cheng},\ and\ \citenamefont {Xu}}]{Yao2011}%
  \BibitemOpen
  \bibfield  {author} {\bibinfo {author} {\bibfnamefont {J.}~\bibnamefont
  {Yao}}, \bibinfo {author} {\bibfnamefont {B.}~\bibnamefont {Zeng}}, \bibinfo
  {author} {\bibfnamefont {H.}~\bibnamefont {Xu}}, \bibinfo {author}
  {\bibfnamefont {G.}~\bibnamefont {Li}}, \bibinfo {author} {\bibfnamefont
  {W.}~\bibnamefont {Chu}}, \bibinfo {author} {\bibfnamefont {J.}~\bibnamefont
  {Ni}}, \bibinfo {author} {\bibfnamefont {H.}~\bibnamefont {Zhang}}, \bibinfo
  {author} {\bibfnamefont {S.~L.}\ \bibnamefont {Chin}}, \bibinfo {author}
  {\bibfnamefont {Y.}~\bibnamefont {Cheng}},\ and\ \bibinfo {author}
  {\bibfnamefont {Z.}~\bibnamefont {Xu}},\ }\bibfield  {title} {\bibinfo
  {title} {High-brightness switchable multiwavelength remote laser in air},\
  }\href {https://doi.org/10.1103/PhysRevA.84.051802} {\bibfield  {journal}
  {\bibinfo  {journal} {Phys. Rev. A}\ }\textbf {\bibinfo {volume} {84}},\
  \bibinfo {pages} {051802} (\bibinfo {year} {2011})}\BibitemShut {NoStop}%
\bibitem [{\citenamefont {Lei}\ \emph {et~al.}(2022)\citenamefont {Lei},
  \citenamefont {Yao}, \citenamefont {Zhao}, \citenamefont {Xie}, \citenamefont
  {Zhang}, \citenamefont {Zhang}, \citenamefont {Zhang}, \citenamefont {Li},
  \citenamefont {Zhang}, \citenamefont {Wang}, \citenamefont {Yang},
  \citenamefont {Yuan}, \citenamefont {Cheng},\ and\ \citenamefont
  {Zhao}}]{zhao2022nc}%
  \BibitemOpen
  \bibfield  {author} {\bibinfo {author} {\bibfnamefont {H.}~\bibnamefont
  {Lei}}, \bibinfo {author} {\bibfnamefont {J.}~\bibnamefont {Yao}}, \bibinfo
  {author} {\bibfnamefont {J.}~\bibnamefont {Zhao}}, \bibinfo {author}
  {\bibfnamefont {H.}~\bibnamefont {Xie}}, \bibinfo {author} {\bibfnamefont
  {F.}~\bibnamefont {Zhang}}, \bibinfo {author} {\bibfnamefont
  {H.}~\bibnamefont {Zhang}}, \bibinfo {author} {\bibfnamefont
  {N.}~\bibnamefont {Zhang}}, \bibinfo {author} {\bibfnamefont
  {G.}~\bibnamefont {Li}}, \bibinfo {author} {\bibfnamefont {Q.}~\bibnamefont
  {Zhang}}, \bibinfo {author} {\bibfnamefont {X.}~\bibnamefont {Wang}},
  \bibinfo {author} {\bibfnamefont {Y.}~\bibnamefont {Yang}}, \bibinfo {author}
  {\bibfnamefont {L.}~\bibnamefont {Yuan}}, \bibinfo {author} {\bibfnamefont
  {Y.}~\bibnamefont {Cheng}},\ and\ \bibinfo {author} {\bibfnamefont
  {Z.}~\bibnamefont {Zhao}},\ }\bibfield  {title} {\bibinfo {title}
  {Ultraviolet supercontinuum generation driven by ionic coherence in a strong
  laser field},\ }\href {https://doi.org/10.1038/s41467-022-31824-0} {\bibfield
   {journal} {\bibinfo  {journal} {Nature Communications}\ }\textbf {\bibinfo
  {volume} {13}},\ \bibinfo {pages} {4080} (\bibinfo {year}
  {2022})}\BibitemShut {NoStop}%
\bibitem [{\citenamefont {Goulielmakis}\ \emph {et~al.}(2010)\citenamefont
  {Goulielmakis}, \citenamefont {Loh}, \citenamefont {Wirth}, \citenamefont
  {Santra}, \citenamefont {Rohringer}, \citenamefont {Yakovlev}, \citenamefont
  {Zherebtsov}, \citenamefont {Pfeifer}, \citenamefont {Azzeer}, \citenamefont
  {Kling}, \citenamefont {Leone},\ and\ \citenamefont
  {Krausz}}]{goulielmakis2010real}%
  \BibitemOpen
  \bibfield  {author} {\bibinfo {author} {\bibfnamefont {E.}~\bibnamefont
  {Goulielmakis}}, \bibinfo {author} {\bibfnamefont {Z.-H.}\ \bibnamefont
  {Loh}}, \bibinfo {author} {\bibfnamefont {A.}~\bibnamefont {Wirth}}, \bibinfo
  {author} {\bibfnamefont {R.}~\bibnamefont {Santra}}, \bibinfo {author}
  {\bibfnamefont {N.}~\bibnamefont {Rohringer}}, \bibinfo {author}
  {\bibfnamefont {V.~S.}\ \bibnamefont {Yakovlev}}, \bibinfo {author}
  {\bibfnamefont {S.}~\bibnamefont {Zherebtsov}}, \bibinfo {author}
  {\bibfnamefont {T.}~\bibnamefont {Pfeifer}}, \bibinfo {author} {\bibfnamefont
  {A.~M.}\ \bibnamefont {Azzeer}}, \bibinfo {author} {\bibfnamefont {M.~F.}\
  \bibnamefont {Kling}}, \bibinfo {author} {\bibfnamefont {S.~R.}\ \bibnamefont
  {Leone}},\ and\ \bibinfo {author} {\bibfnamefont {F.}~\bibnamefont
  {Krausz}},\ }\bibfield  {title} {\bibinfo {title} {Real-time observation of
  valence electron motion},\ }\href {https://doi.org/10.1038/nature09212}
  {\bibfield  {journal} {\bibinfo  {journal} {Nature}\ }\textbf {\bibinfo
  {volume} {466}},\ \bibinfo {pages} {739} (\bibinfo {year}
  {2010})}\BibitemShut {NoStop}%
\bibitem [{\citenamefont {Sabbar}\ \emph {et~al.}(2017)\citenamefont {Sabbar},
  \citenamefont {Timmers}, \citenamefont {Chen}, \citenamefont {Pymer},
  \citenamefont {Loh}, \citenamefont {Sayres}, \citenamefont {Pabst},
  \citenamefont {Santra},\ and\ \citenamefont {Leone}}]{sabbar2017np}%
  \BibitemOpen
  \bibfield  {author} {\bibinfo {author} {\bibfnamefont {M.}~\bibnamefont
  {Sabbar}}, \bibinfo {author} {\bibfnamefont {H.}~\bibnamefont {Timmers}},
  \bibinfo {author} {\bibfnamefont {Y.-J.}\ \bibnamefont {Chen}}, \bibinfo
  {author} {\bibfnamefont {A.~K.}\ \bibnamefont {Pymer}}, \bibinfo {author}
  {\bibfnamefont {Z.-H.}\ \bibnamefont {Loh}}, \bibinfo {author} {\bibfnamefont
  {S.~G.}\ \bibnamefont {Sayres}}, \bibinfo {author} {\bibfnamefont
  {S.}~\bibnamefont {Pabst}}, \bibinfo {author} {\bibfnamefont
  {R.}~\bibnamefont {Santra}},\ and\ \bibinfo {author} {\bibfnamefont {S.~R.}\
  \bibnamefont {Leone}},\ }\bibfield  {title} {\bibinfo {title} {State-resolved
  attosecond reversible and irreversible dynamics in strong optical fields},\
  }\href {https://doi.org/10.1038/nphys4027} {\bibfield  {journal} {\bibinfo
  {journal} {Nature Physics}\ }\textbf {\bibinfo {volume} {13}},\ \bibinfo
  {pages} {472} (\bibinfo {year} {2017})}\BibitemShut {NoStop}%
\bibitem [{\citenamefont {Saito}\ \emph {et~al.}(2019)\citenamefont {Saito},
  \citenamefont {Sannohe}, \citenamefont {Ishii}, \citenamefont {Kanai},
  \citenamefont {Kosugi}, \citenamefont {Wu}, \citenamefont {Chew},
  \citenamefont {Han}, \citenamefont {Chang},\ and\ \citenamefont
  {Itatani}}]{saito2019real}%
  \BibitemOpen
  \bibfield  {author} {\bibinfo {author} {\bibfnamefont {N.}~\bibnamefont
  {Saito}}, \bibinfo {author} {\bibfnamefont {H.}~\bibnamefont {Sannohe}},
  \bibinfo {author} {\bibfnamefont {N.}~\bibnamefont {Ishii}}, \bibinfo
  {author} {\bibfnamefont {T.}~\bibnamefont {Kanai}}, \bibinfo {author}
  {\bibfnamefont {N.}~\bibnamefont {Kosugi}}, \bibinfo {author} {\bibfnamefont
  {Y.}~\bibnamefont {Wu}}, \bibinfo {author} {\bibfnamefont {A.}~\bibnamefont
  {Chew}}, \bibinfo {author} {\bibfnamefont {S.}~\bibnamefont {Han}}, \bibinfo
  {author} {\bibfnamefont {Z.}~\bibnamefont {Chang}},\ and\ \bibinfo {author}
  {\bibfnamefont {J.}~\bibnamefont {Itatani}},\ }\bibfield  {title} {\bibinfo
  {title} {Real-time observation of electronic, vibrational, and rotational
  dynamics in nitric oxide with attosecond soft x-ray pulses at 400 e{V}},\
  }\href {https://doi.org/10.1364/OPTICA.6.001542} {\bibfield  {journal}
  {\bibinfo  {journal} {Optica}\ }\textbf {\bibinfo {volume} {6}},\ \bibinfo
  {pages} {1542} (\bibinfo {year} {2019})}\BibitemShut {NoStop}%
\bibitem [{\citenamefont {Smirnova}\ \emph {et~al.}(2009)\citenamefont
  {Smirnova}, \citenamefont {Patchkovskii}, \citenamefont {Mairesse},
  \citenamefont {Dudovich},\ and\ \citenamefont {Ivanov}}]{Smirnova09S}%
  \BibitemOpen
  \bibfield  {author} {\bibinfo {author} {\bibfnamefont {O.}~\bibnamefont
  {Smirnova}}, \bibinfo {author} {\bibfnamefont {S.}~\bibnamefont
  {Patchkovskii}}, \bibinfo {author} {\bibfnamefont {Y.}~\bibnamefont
  {Mairesse}}, \bibinfo {author} {\bibfnamefont {N.}~\bibnamefont {Dudovich}},\
  and\ \bibinfo {author} {\bibfnamefont {M.~Y.}\ \bibnamefont {Ivanov}},\
  }\bibfield  {title} {\bibinfo {title} {Strong-field control and spectroscopy
  of attosecond electron-hole dynamics in molecules},\ }\href
  {https://doi.org/10.1073/pnas.0907434106} {\bibfield  {journal} {\bibinfo
  {journal} {Proceedings of the National Academy of Sciences}\ }\textbf
  {\bibinfo {volume} {106}},\ \bibinfo {pages} {16556} (\bibinfo {year}
  {2009})}\BibitemShut {NoStop}%
\bibitem [{\citenamefont {Rohringer}\ and\ \citenamefont
  {Santra}(2009)}]{Rohringer2009}%
  \BibitemOpen
  \bibfield  {author} {\bibinfo {author} {\bibfnamefont {N.}~\bibnamefont
  {Rohringer}}\ and\ \bibinfo {author} {\bibfnamefont {R.}~\bibnamefont
  {Santra}},\ }\bibfield  {title} {\bibinfo {title} {Multichannel coherence in
  strong-field ionization},\ }\href
  {https://doi.org/10.1103/PhysRevA.79.053402} {\bibfield  {journal} {\bibinfo
  {journal} {Phys. Rev. A}\ }\textbf {\bibinfo {volume} {79}},\ \bibinfo
  {pages} {053402} (\bibinfo {year} {2009})}\BibitemShut {NoStop}%
\bibitem [{\citenamefont {Pabst}\ \emph {et~al.}(2012)\citenamefont {Pabst},
  \citenamefont {Sytcheva}, \citenamefont {Moulet}, \citenamefont {Wirth},
  \citenamefont {Goulielmakis},\ and\ \citenamefont {Santra}}]{Pabst2012A2}%
  \BibitemOpen
  \bibfield  {author} {\bibinfo {author} {\bibfnamefont {S.}~\bibnamefont
  {Pabst}}, \bibinfo {author} {\bibfnamefont {A.}~\bibnamefont {Sytcheva}},
  \bibinfo {author} {\bibfnamefont {A.}~\bibnamefont {Moulet}}, \bibinfo
  {author} {\bibfnamefont {A.}~\bibnamefont {Wirth}}, \bibinfo {author}
  {\bibfnamefont {E.}~\bibnamefont {Goulielmakis}},\ and\ \bibinfo {author}
  {\bibfnamefont {R.}~\bibnamefont {Santra}},\ }\bibfield  {title} {\bibinfo
  {title} {Theory of attosecond transient-absorption spectroscopy of krypton
  for overlapping pump and probe pulses},\ }\href
  {https://doi.org/10.1103/PhysRevA.86.063411} {\bibfield  {journal} {\bibinfo
  {journal} {Phys. Rev. A}\ }\textbf {\bibinfo {volume} {86}},\ \bibinfo
  {pages} {063411} (\bibinfo {year} {2012})}\BibitemShut {NoStop}%
\bibitem [{\citenamefont {Zhang}\ \emph {et~al.}(2020)\citenamefont {Zhang},
  \citenamefont {Xie}, \citenamefont {Li}, \citenamefont {Wang}, \citenamefont
  {Lei}, \citenamefont {Zhao}, \citenamefont {Chen}, \citenamefont {Yao},
  \citenamefont {Cheng},\ and\ \citenamefont {Zhao}}]{zhang2020sub}%
  \BibitemOpen
  \bibfield  {author} {\bibinfo {author} {\bibfnamefont {Q.}~\bibnamefont
  {Zhang}}, \bibinfo {author} {\bibfnamefont {H.}~\bibnamefont {Xie}}, \bibinfo
  {author} {\bibfnamefont {G.}~\bibnamefont {Li}}, \bibinfo {author}
  {\bibfnamefont {X.}~\bibnamefont {Wang}}, \bibinfo {author} {\bibfnamefont
  {H.}~\bibnamefont {Lei}}, \bibinfo {author} {\bibfnamefont {J.}~\bibnamefont
  {Zhao}}, \bibinfo {author} {\bibfnamefont {Z.}~\bibnamefont {Chen}}, \bibinfo
  {author} {\bibfnamefont {J.}~\bibnamefont {Yao}}, \bibinfo {author}
  {\bibfnamefont {Y.}~\bibnamefont {Cheng}},\ and\ \bibinfo {author}
  {\bibfnamefont {Z.}~\bibnamefont {Zhao}},\ }\bibfield  {title} {\bibinfo
  {title} {Sub-cycle coherent control of ionic dynamics via transient
  ionization injection},\ }\href {https://doi.org/10.1038/s42005-020-0321-7}
  {\bibfield  {journal} {\bibinfo  {journal} {Communications Physics}\ }\textbf
  {\bibinfo {volume} {3}},\ \bibinfo {pages} {50} (\bibinfo {year}
  {2020})}\BibitemShut {NoStop}%
\bibitem [{\citenamefont {Wang}\ \emph {et~al.}(2024)\citenamefont {Wang},
  \citenamefont {Bai}, \citenamefont {Wang}, \citenamefont {Zhao},
  \citenamefont {Gao}, \citenamefont {Wang}, \citenamefont {Xiao},
  \citenamefont {Tao}, \citenamefont {Song}, \citenamefont {Qiu}, \citenamefont
  {Liu},\ and\ \citenamefont {Zhao}}]{zhao2024nc}%
  \BibitemOpen
  \bibfield  {author} {\bibinfo {author} {\bibfnamefont {L.}~\bibnamefont
  {Wang}}, \bibinfo {author} {\bibfnamefont {G.}~\bibnamefont {Bai}}, \bibinfo
  {author} {\bibfnamefont {X.}~\bibnamefont {Wang}}, \bibinfo {author}
  {\bibfnamefont {J.}~\bibnamefont {Zhao}}, \bibinfo {author} {\bibfnamefont
  {C.}~\bibnamefont {Gao}}, \bibinfo {author} {\bibfnamefont {J.}~\bibnamefont
  {Wang}}, \bibinfo {author} {\bibfnamefont {F.}~\bibnamefont {Xiao}}, \bibinfo
  {author} {\bibfnamefont {W.}~\bibnamefont {Tao}}, \bibinfo {author}
  {\bibfnamefont {P.}~\bibnamefont {Song}}, \bibinfo {author} {\bibfnamefont
  {Q.}~\bibnamefont {Qiu}}, \bibinfo {author} {\bibfnamefont {J.}~\bibnamefont
  {Liu}},\ and\ \bibinfo {author} {\bibfnamefont {Z.}~\bibnamefont {Zhao}},\
  }\bibfield  {title} {\bibinfo {title} {Raman time-delay in attosecond
  transient absorption of strong-field created krypton vacancy},\ }\href
  {https://doi.org/10.1038/s41467-024-47088-9} {\bibfield  {journal} {\bibinfo
  {journal} {Nature Communications}\ }\textbf {\bibinfo {volume} {15}},\
  \bibinfo {pages} {2705} (\bibinfo {year} {2024})}\BibitemShut {NoStop}%
\bibitem [{\citenamefont {Yao}\ \emph {et~al.}(2016)\citenamefont {Yao},
  \citenamefont {Jiang}, \citenamefont {Chu}, \citenamefont {Zeng},
  \citenamefont {Wu}, \citenamefont {Lu}, \citenamefont {Li}, \citenamefont
  {Xie}, \citenamefont {Li}, \citenamefont {Yu}, \citenamefont {Wang},
  \citenamefont {Jiang}, \citenamefont {Gong},\ and\ \citenamefont
  {Cheng}}]{yao2016population}%
  \BibitemOpen
  \bibfield  {author} {\bibinfo {author} {\bibfnamefont {J.}~\bibnamefont
  {Yao}}, \bibinfo {author} {\bibfnamefont {S.}~\bibnamefont {Jiang}}, \bibinfo
  {author} {\bibfnamefont {W.}~\bibnamefont {Chu}}, \bibinfo {author}
  {\bibfnamefont {B.}~\bibnamefont {Zeng}}, \bibinfo {author} {\bibfnamefont
  {C.}~\bibnamefont {Wu}}, \bibinfo {author} {\bibfnamefont {R.}~\bibnamefont
  {Lu}}, \bibinfo {author} {\bibfnamefont {Z.}~\bibnamefont {Li}}, \bibinfo
  {author} {\bibfnamefont {H.}~\bibnamefont {Xie}}, \bibinfo {author}
  {\bibfnamefont {G.}~\bibnamefont {Li}}, \bibinfo {author} {\bibfnamefont
  {C.}~\bibnamefont {Yu}}, \bibinfo {author} {\bibfnamefont {Z.}~\bibnamefont
  {Wang}}, \bibinfo {author} {\bibfnamefont {H.}~\bibnamefont {Jiang}},
  \bibinfo {author} {\bibfnamefont {Q.}~\bibnamefont {Gong}},\ and\ \bibinfo
  {author} {\bibfnamefont {Y.}~\bibnamefont {Cheng}},\ }\bibfield  {title}
  {\bibinfo {title} {Population redistribution among multiple electronic states
  of molecular nitrogen ions in strong laser fields},\ }\href
  {https://doi.org/10.1103/PhysRevLett.116.143007} {\bibfield  {journal}
  {\bibinfo  {journal} {Phys. Rev. Lett.}\ }\textbf {\bibinfo {volume} {116}},\
  \bibinfo {pages} {143007} (\bibinfo {year} {2016})}\BibitemShut {NoStop}%
\bibitem [{\citenamefont {Xu}\ \emph {et~al.}(2015)\citenamefont {Xu},
  \citenamefont {L{\"o}tstedt}, \citenamefont {Iwasaki},\ and\ \citenamefont
  {Yamanouchi}}]{xu2015sub}%
  \BibitemOpen
  \bibfield  {author} {\bibinfo {author} {\bibfnamefont {H.}~\bibnamefont
  {Xu}}, \bibinfo {author} {\bibfnamefont {E.}~\bibnamefont {L{\"o}tstedt}},
  \bibinfo {author} {\bibfnamefont {A.}~\bibnamefont {Iwasaki}},\ and\ \bibinfo
  {author} {\bibfnamefont {K.}~\bibnamefont {Yamanouchi}},\ }\bibfield  {title}
  {\bibinfo {title} {Sub-10-fs population inversion in n2+ in air lasing
  through multiple state coupling},\ }\href
  {https://doi.org/10.1038/ncomms9347} {\bibfield  {journal} {\bibinfo
  {journal} {Nature Communications}\ }\textbf {\bibinfo {volume} {6}},\
  \bibinfo {pages} {8347} (\bibinfo {year} {2015})}\BibitemShut {NoStop}%
\bibitem [{\citenamefont {Liu}\ \emph {et~al.}(2015)\citenamefont {Liu},
  \citenamefont {Ding}, \citenamefont {Lambert}, \citenamefont {Houard},
  \citenamefont {Tikhonchuk},\ and\ \citenamefont {Mysyrowicz}}]{liu15prl}%
  \BibitemOpen
  \bibfield  {author} {\bibinfo {author} {\bibfnamefont {Y.}~\bibnamefont
  {Liu}}, \bibinfo {author} {\bibfnamefont {P.}~\bibnamefont {Ding}}, \bibinfo
  {author} {\bibfnamefont {G.}~\bibnamefont {Lambert}}, \bibinfo {author}
  {\bibfnamefont {A.}~\bibnamefont {Houard}}, \bibinfo {author} {\bibfnamefont
  {V.}~\bibnamefont {Tikhonchuk}},\ and\ \bibinfo {author} {\bibfnamefont
  {A.}~\bibnamefont {Mysyrowicz}},\ }\bibfield  {title} {\bibinfo {title}
  {Recollision-induced superradiance of ionized nitrogen molecules},\ }\href
  {https://doi.org/10.1103/PhysRevLett.115.133203} {\bibfield  {journal}
  {\bibinfo  {journal} {Phys Rev Lett}\ }\textbf {\bibinfo {volume} {115}},\
  \bibinfo {pages} {133203} (\bibinfo {year} {2015})}\BibitemShut {NoStop}%
\bibitem [{\citenamefont {Britton}\ \emph {et~al.}(2018)\citenamefont
  {Britton}, \citenamefont {Laferri\`ere}, \citenamefont {Ko}, \citenamefont
  {Li}, \citenamefont {Kong}, \citenamefont {Brown}, \citenamefont {Naumov},
  \citenamefont {Zhang}, \citenamefont {Arissian},\ and\ \citenamefont
  {Corkum}}]{corkum18prl}%
  \BibitemOpen
  \bibfield  {author} {\bibinfo {author} {\bibfnamefont {M.}~\bibnamefont
  {Britton}}, \bibinfo {author} {\bibfnamefont {P.}~\bibnamefont
  {Laferri\`ere}}, \bibinfo {author} {\bibfnamefont {D.~H.}\ \bibnamefont
  {Ko}}, \bibinfo {author} {\bibfnamefont {Z.}~\bibnamefont {Li}}, \bibinfo
  {author} {\bibfnamefont {F.}~\bibnamefont {Kong}}, \bibinfo {author}
  {\bibfnamefont {G.}~\bibnamefont {Brown}}, \bibinfo {author} {\bibfnamefont
  {A.}~\bibnamefont {Naumov}}, \bibinfo {author} {\bibfnamefont
  {C.}~\bibnamefont {Zhang}}, \bibinfo {author} {\bibfnamefont
  {L.}~\bibnamefont {Arissian}},\ and\ \bibinfo {author} {\bibfnamefont
  {P.~B.}\ \bibnamefont {Corkum}},\ }\bibfield  {title} {\bibinfo {title}
  {Testing the role of recollision in ${\mathrm{n}}_{2}^{+}$ air lasing},\
  }\href {https://doi.org/10.1103/PhysRevLett.120.133208} {\bibfield  {journal}
  {\bibinfo  {journal} {Phys. Rev. Lett.}\ }\textbf {\bibinfo {volume} {120}},\
  \bibinfo {pages} {133208} (\bibinfo {year} {2018})}\BibitemShut {NoStop}%
\bibitem [{\citenamefont {Li}\ \emph {et~al.}(2020)\citenamefont {Li},
  \citenamefont {L\"otstedt}, \citenamefont {Li}, \citenamefont {Zhou},
  \citenamefont {Dong}, \citenamefont {Deng}, \citenamefont {Lu}, \citenamefont
  {Ando}, \citenamefont {Iwasaki}, \citenamefont {Fu}, \citenamefont {Wang},
  \citenamefont {Wu}, \citenamefont {Yamanouchi},\ and\ \citenamefont
  {Xu}}]{xu2020prl}%
  \BibitemOpen
  \bibfield  {author} {\bibinfo {author} {\bibfnamefont {H.}~\bibnamefont
  {Li}}, \bibinfo {author} {\bibfnamefont {E.}~\bibnamefont {L\"otstedt}},
  \bibinfo {author} {\bibfnamefont {H.}~\bibnamefont {Li}}, \bibinfo {author}
  {\bibfnamefont {Y.}~\bibnamefont {Zhou}}, \bibinfo {author} {\bibfnamefont
  {N.}~\bibnamefont {Dong}}, \bibinfo {author} {\bibfnamefont {L.}~\bibnamefont
  {Deng}}, \bibinfo {author} {\bibfnamefont {P.}~\bibnamefont {Lu}}, \bibinfo
  {author} {\bibfnamefont {T.}~\bibnamefont {Ando}}, \bibinfo {author}
  {\bibfnamefont {A.}~\bibnamefont {Iwasaki}}, \bibinfo {author} {\bibfnamefont
  {Y.}~\bibnamefont {Fu}}, \bibinfo {author} {\bibfnamefont {S.}~\bibnamefont
  {Wang}}, \bibinfo {author} {\bibfnamefont {J.}~\bibnamefont {Wu}}, \bibinfo
  {author} {\bibfnamefont {K.}~\bibnamefont {Yamanouchi}},\ and\ \bibinfo
  {author} {\bibfnamefont {H.}~\bibnamefont {Xu}},\ }\bibfield  {title}
  {\bibinfo {title} {Giant enhancement of air lasing by complete population
  inversion in ${\mathrm{n}}_{2}^{+}$},\ }\href
  {https://doi.org/10.1103/PhysRevLett.125.053201} {\bibfield  {journal}
  {\bibinfo  {journal} {Phys. Rev. Lett.}\ }\textbf {\bibinfo {volume} {125}},\
  \bibinfo {pages} {053201} (\bibinfo {year} {2020})}\BibitemShut {NoStop}%
\bibitem [{\citenamefont {Mysyrowicz}\ \emph {et~al.}(2019)\citenamefont
  {Mysyrowicz}, \citenamefont {Danylo}, \citenamefont {Houard}, \citenamefont
  {Tikhonchuk}, \citenamefont {Zhang}, \citenamefont {Fan}, \citenamefont
  {Liang}, \citenamefont {Zhuang}, \citenamefont {Yuan},\ and\ \citenamefont
  {Liu}}]{Mysyrowicz2019}%
  \BibitemOpen
  \bibfield  {author} {\bibinfo {author} {\bibfnamefont {A.}~\bibnamefont
  {Mysyrowicz}}, \bibinfo {author} {\bibfnamefont {R.}~\bibnamefont {Danylo}},
  \bibinfo {author} {\bibfnamefont {A.}~\bibnamefont {Houard}}, \bibinfo
  {author} {\bibfnamefont {V.}~\bibnamefont {Tikhonchuk}}, \bibinfo {author}
  {\bibfnamefont {X.}~\bibnamefont {Zhang}}, \bibinfo {author} {\bibfnamefont
  {Z.}~\bibnamefont {Fan}}, \bibinfo {author} {\bibfnamefont {Q.}~\bibnamefont
  {Liang}}, \bibinfo {author} {\bibfnamefont {S.}~\bibnamefont {Zhuang}},
  \bibinfo {author} {\bibfnamefont {L.}~\bibnamefont {Yuan}},\ and\ \bibinfo
  {author} {\bibfnamefont {Y.}~\bibnamefont {Liu}},\ }\bibfield  {title}
  {\bibinfo {title} {Lasing without population inversion in
  ${\mathrm{n}}_{2}^{+}$},\ }\href {https://doi.org/10.1063/1.5116898}
  {\bibfield  {journal} {\bibinfo  {journal} {APL Photonics}\ }\textbf
  {\bibinfo {volume} {4}},\ \bibinfo {pages} {110807} (\bibinfo {year}
  {2019})}\BibitemShut {NoStop}%
\bibitem [{\citenamefont {Arissian}\ \emph {et~al.}(2018)\citenamefont
  {Arissian}, \citenamefont {Kamer}, \citenamefont {Rastegari}, \citenamefont
  {Villeneuve},\ and\ \citenamefont {Diels}}]{arissian2018pra}%
  \BibitemOpen
  \bibfield  {author} {\bibinfo {author} {\bibfnamefont {L.}~\bibnamefont
  {Arissian}}, \bibinfo {author} {\bibfnamefont {B.}~\bibnamefont {Kamer}},
  \bibinfo {author} {\bibfnamefont {A.}~\bibnamefont {Rastegari}}, \bibinfo
  {author} {\bibfnamefont {D.~M.}\ \bibnamefont {Villeneuve}},\ and\ \bibinfo
  {author} {\bibfnamefont {J.-C.}\ \bibnamefont {Diels}},\ }\bibfield  {title}
  {\bibinfo {title} {Transient gain from ${{\mathrm{N}}_{2}}^{+}$ in light
  filaments},\ }\href {https://doi.org/10.1103/PhysRevA.98.053438} {\bibfield
  {journal} {\bibinfo  {journal} {Phys. Rev. A}\ }\textbf {\bibinfo {volume}
  {98}},\ \bibinfo {pages} {053438} (\bibinfo {year} {2018})}\BibitemShut
  {NoStop}%
\bibitem [{\citenamefont {Richter}\ \emph {et~al.}(2020)\citenamefont
  {Richter}, \citenamefont {Lytova}, \citenamefont {Morales}, \citenamefont
  {Haessler}, \citenamefont {Smirnova}, \citenamefont {Spanner},\ and\
  \citenamefont {Ivanov}}]{richter2020rotational}%
  \BibitemOpen
  \bibfield  {author} {\bibinfo {author} {\bibfnamefont {M.}~\bibnamefont
  {Richter}}, \bibinfo {author} {\bibfnamefont {M.}~\bibnamefont {Lytova}},
  \bibinfo {author} {\bibfnamefont {F.}~\bibnamefont {Morales}}, \bibinfo
  {author} {\bibfnamefont {S.}~\bibnamefont {Haessler}}, \bibinfo {author}
  {\bibfnamefont {O.}~\bibnamefont {Smirnova}}, \bibinfo {author}
  {\bibfnamefont {M.}~\bibnamefont {Spanner}},\ and\ \bibinfo {author}
  {\bibfnamefont {M.}~\bibnamefont {Ivanov}},\ }\bibfield  {title} {\bibinfo
  {title} {Rotational quantum beat lasing without inversion},\ }\href
  {https://doi.org/10.1364/OPTICA.390665} {\bibfield  {journal} {\bibinfo
  {journal} {Optica}\ }\textbf {\bibinfo {volume} {7}},\ \bibinfo {pages} {586}
  (\bibinfo {year} {2020})}\BibitemShut {NoStop}%
\bibitem [{\citenamefont {Xie}\ \emph {et~al.}(2020)\citenamefont {Xie},
  \citenamefont {Lei}, \citenamefont {Li}, \citenamefont {Zhang}, \citenamefont
  {Wang}, \citenamefont {Zhao}, \citenamefont {Chen}, \citenamefont {Yao},
  \citenamefont {Cheng},\ and\ \citenamefont {Zhao}}]{xie2020role}%
  \BibitemOpen
  \bibfield  {author} {\bibinfo {author} {\bibfnamefont {H.}~\bibnamefont
  {Xie}}, \bibinfo {author} {\bibfnamefont {H.}~\bibnamefont {Lei}}, \bibinfo
  {author} {\bibfnamefont {G.}~\bibnamefont {Li}}, \bibinfo {author}
  {\bibfnamefont {Q.}~\bibnamefont {Zhang}}, \bibinfo {author} {\bibfnamefont
  {X.}~\bibnamefont {Wang}}, \bibinfo {author} {\bibfnamefont {J.}~\bibnamefont
  {Zhao}}, \bibinfo {author} {\bibfnamefont {Z.}~\bibnamefont {Chen}}, \bibinfo
  {author} {\bibfnamefont {J.}~\bibnamefont {Yao}}, \bibinfo {author}
  {\bibfnamefont {Y.}~\bibnamefont {Cheng}},\ and\ \bibinfo {author}
  {\bibfnamefont {Z.}~\bibnamefont {Zhao}},\ }\bibfield  {title} {\bibinfo
  {title} {Role of rotational coherence in femtosecond-pulse-driven nitrogen
  ion lasing},\ }\href {https://doi.org/10.1103/PhysRevResearch.2.023329}
  {\bibfield  {journal} {\bibinfo  {journal} {Phys. Rev. Res.}\ }\textbf
  {\bibinfo {volume} {2}},\ \bibinfo {pages} {023329} (\bibinfo {year}
  {2020})}\BibitemShut {NoStop}%
\bibitem [{\citenamefont {Zhang}\ \emph {et~al.}(2021)\citenamefont {Zhang},
  \citenamefont {L\"otstedt}, \citenamefont {Ando}, \citenamefont {Iwasaki},
  \citenamefont {Xu},\ and\ \citenamefont {Yamanouchi}}]{erik2021pra}%
  \BibitemOpen
  \bibfield  {author} {\bibinfo {author} {\bibfnamefont {Y.}~\bibnamefont
  {Zhang}}, \bibinfo {author} {\bibfnamefont {E.}~\bibnamefont {L\"otstedt}},
  \bibinfo {author} {\bibfnamefont {T.}~\bibnamefont {Ando}}, \bibinfo {author}
  {\bibfnamefont {A.}~\bibnamefont {Iwasaki}}, \bibinfo {author} {\bibfnamefont
  {H.}~\bibnamefont {Xu}},\ and\ \bibinfo {author} {\bibfnamefont
  {K.}~\bibnamefont {Yamanouchi}},\ }\bibfield  {title} {\bibinfo {title}
  {Rotational population transfer through the
  $a{\phantom{\rule{0.16em}{0ex}}}^{2}{\mathrm{\ensuremath{\Pi}}}_{u}\ensuremath{-}x{\phantom{\rule{0.16em}{0ex}}}^{2}{\mathrm{\ensuremath{\Sigma}}}_{g}{}^{+}\ensuremath{-}b{\phantom{\rule{0.16em}{0ex}}}^{2}{\mathrm{\ensuremath{\Sigma}}}_{u}{}^{+}$
  coupling in ${\mathrm{n}}_{2}{}^{+}$ lasing},\ }\href
  {https://doi.org/10.1103/PhysRevA.104.023107} {\bibfield  {journal} {\bibinfo
   {journal} {Phys. Rev. A}\ }\textbf {\bibinfo {volume} {104}},\ \bibinfo
  {pages} {023107} (\bibinfo {year} {2021})}\BibitemShut {NoStop}%
\bibitem [{\citenamefont {Ando}\ \emph {et~al.}(2019)\citenamefont {Ando},
  \citenamefont {L\"otstedt}, \citenamefont {Iwasaki}, \citenamefont {Li},
  \citenamefont {Fu}, \citenamefont {Wang}, \citenamefont {Xu},\ and\
  \citenamefont {Yamanouchi}}]{Ando2019Rota}%
  \BibitemOpen
  \bibfield  {author} {\bibinfo {author} {\bibfnamefont {T.}~\bibnamefont
  {Ando}}, \bibinfo {author} {\bibfnamefont {E.}~\bibnamefont {L\"otstedt}},
  \bibinfo {author} {\bibfnamefont {A.}~\bibnamefont {Iwasaki}}, \bibinfo
  {author} {\bibfnamefont {H.}~\bibnamefont {Li}}, \bibinfo {author}
  {\bibfnamefont {Y.}~\bibnamefont {Fu}}, \bibinfo {author} {\bibfnamefont
  {S.}~\bibnamefont {Wang}}, \bibinfo {author} {\bibfnamefont {H.}~\bibnamefont
  {Xu}},\ and\ \bibinfo {author} {\bibfnamefont {K.}~\bibnamefont
  {Yamanouchi}},\ }\bibfield  {title} {\bibinfo {title} {Rotational,
  vibrational, and electronic modulations in {N}$_2^+$ lasing at 391 nm:
  Evidence of coherent
  $b{^{2}\mathrm{\ensuremath{\Sigma}}}_{u}^{+}\ensuremath{-}x{^{2}\mathrm{\ensuremath{\Sigma}}}_{g}^{+}\ensuremath{-}a{^{2}\mathrm{\ensuremath{\Pi}}}_{u}$
  coupling},\ }\href {https://doi.org/10.1103/PhysRevLett.123.203201}
  {\bibfield  {journal} {\bibinfo  {journal} {Phys. Rev. Lett.}\ }\textbf
  {\bibinfo {volume} {123}},\ \bibinfo {pages} {203201} (\bibinfo {year}
  {2019})}\BibitemShut {NoStop}%
\bibitem [{\citenamefont {Tikhonchuk}\ \emph {et~al.}(2021)\citenamefont
  {Tikhonchuk}, \citenamefont {Liu}, \citenamefont {Danylo}, \citenamefont
  {Houard},\ and\ \citenamefont {Mysyrowicz}}]{Tikhonchuk2021NJ}%
  \BibitemOpen
  \bibfield  {author} {\bibinfo {author} {\bibfnamefont {V.~T.}\ \bibnamefont
  {Tikhonchuk}}, \bibinfo {author} {\bibfnamefont {Y.}~\bibnamefont {Liu}},
  \bibinfo {author} {\bibfnamefont {R.}~\bibnamefont {Danylo}}, \bibinfo
  {author} {\bibfnamefont {A.}~\bibnamefont {Houard}},\ and\ \bibinfo {author}
  {\bibfnamefont {A.}~\bibnamefont {Mysyrowicz}},\ }\bibfield  {title}
  {\bibinfo {title} {Theory of femtosecond strong field ion excitation and
  subsequent lasing in {N}$_2^+$},\ }\href
  {https://doi.org/10.1088/1367-2630/abd8bf} {\bibfield  {journal} {\bibinfo
  {journal} {New Journal of Physics}\ }\textbf {\bibinfo {volume} {23}},\
  \bibinfo {pages} {023035} (\bibinfo {year} {2021})}\BibitemShut {NoStop}%
\bibitem [{\citenamefont {Golubev}\ \emph {et~al.}(2021)\citenamefont
  {Golubev}, \citenamefont {Van\'{\i}\ifmmode~\check{c}\else \v{c}\fi{}ek},\
  and\ \citenamefont {Kuleff}}]{golubev2021prl}%
  \BibitemOpen
  \bibfield  {author} {\bibinfo {author} {\bibfnamefont {N.~V.}\ \bibnamefont
  {Golubev}}, \bibinfo {author} {\bibfnamefont {J.}~\bibnamefont
  {Van\'{\i}\ifmmode~\check{c}\else \v{c}\fi{}ek}},\ and\ \bibinfo {author}
  {\bibfnamefont {A.~I.}\ \bibnamefont {Kuleff}},\ }\bibfield  {title}
  {\bibinfo {title} {Core-valence attosecond transient absorption spectroscopy
  of polyatomic molecules},\ }\href
  {https://doi.org/10.1103/PhysRevLett.127.123001} {\bibfield  {journal}
  {\bibinfo  {journal} {Phys. Rev. Lett.}\ }\textbf {\bibinfo {volume} {127}},\
  \bibinfo {pages} {123001} (\bibinfo {year} {2021})}\BibitemShut {NoStop}%
\bibitem [{\citenamefont {Magunia}\ \emph {et~al.}(2023)\citenamefont
  {Magunia}, \citenamefont {Rebholz}, \citenamefont {Appi}, \citenamefont
  {Papadopoulou}, \citenamefont {Lindenblatt}, \citenamefont {Trost},
  \citenamefont {Meister}, \citenamefont {Ding}, \citenamefont {Straub},
  \citenamefont {Borisova}, \citenamefont {Lee}, \citenamefont {Jin},
  \citenamefont {von~der Dellen}, \citenamefont {Kaiser}, \citenamefont
  {Braune}, \citenamefont {Düsterer}, \citenamefont {Ališauskas},
  \citenamefont {Lang}, \citenamefont {Heyl}, \citenamefont {Manschwetus},
  \citenamefont {Grunewald}, \citenamefont {Frühling}, \citenamefont
  {Tajalli}, \citenamefont {Wahid}, \citenamefont {Silletti}, \citenamefont
  {Calegari}, \citenamefont {Mosel}, \citenamefont {Morgner}, \citenamefont
  {Kovacev}, \citenamefont {Thumm}, \citenamefont {Hartl}, \citenamefont
  {Treusch}, \citenamefont {Moshammer}, \citenamefont {Ott},\ and\
  \citenamefont {Pfeifer}}]{pfeifer2023sa}%
  \BibitemOpen
  \bibfield  {author} {\bibinfo {author} {\bibfnamefont {A.}~\bibnamefont
  {Magunia}}, \bibinfo {author} {\bibfnamefont {M.}~\bibnamefont {Rebholz}},
  \bibinfo {author} {\bibfnamefont {E.}~\bibnamefont {Appi}}, \bibinfo {author}
  {\bibfnamefont {C.~C.}\ \bibnamefont {Papadopoulou}}, \bibinfo {author}
  {\bibfnamefont {H.}~\bibnamefont {Lindenblatt}}, \bibinfo {author}
  {\bibfnamefont {F.}~\bibnamefont {Trost}}, \bibinfo {author} {\bibfnamefont
  {S.}~\bibnamefont {Meister}}, \bibinfo {author} {\bibfnamefont
  {T.}~\bibnamefont {Ding}}, \bibinfo {author} {\bibfnamefont {M.}~\bibnamefont
  {Straub}}, \bibinfo {author} {\bibfnamefont {G.~D.}\ \bibnamefont
  {Borisova}}, \bibinfo {author} {\bibfnamefont {J.}~\bibnamefont {Lee}},
  \bibinfo {author} {\bibfnamefont {R.}~\bibnamefont {Jin}}, \bibinfo {author}
  {\bibfnamefont {A.}~\bibnamefont {von~der Dellen}}, \bibinfo {author}
  {\bibfnamefont {C.}~\bibnamefont {Kaiser}}, \bibinfo {author} {\bibfnamefont
  {M.}~\bibnamefont {Braune}}, \bibinfo {author} {\bibfnamefont
  {S.}~\bibnamefont {Düsterer}}, \bibinfo {author} {\bibfnamefont
  {S.}~\bibnamefont {Ališauskas}}, \bibinfo {author} {\bibfnamefont
  {T.}~\bibnamefont {Lang}}, \bibinfo {author} {\bibfnamefont {C.}~\bibnamefont
  {Heyl}}, \bibinfo {author} {\bibfnamefont {B.}~\bibnamefont {Manschwetus}},
  \bibinfo {author} {\bibfnamefont {S.}~\bibnamefont {Grunewald}}, \bibinfo
  {author} {\bibfnamefont {U.}~\bibnamefont {Frühling}}, \bibinfo {author}
  {\bibfnamefont {A.}~\bibnamefont {Tajalli}}, \bibinfo {author} {\bibfnamefont
  {A.~B.}\ \bibnamefont {Wahid}}, \bibinfo {author} {\bibfnamefont
  {L.}~\bibnamefont {Silletti}}, \bibinfo {author} {\bibfnamefont
  {F.}~\bibnamefont {Calegari}}, \bibinfo {author} {\bibfnamefont
  {P.}~\bibnamefont {Mosel}}, \bibinfo {author} {\bibfnamefont
  {U.}~\bibnamefont {Morgner}}, \bibinfo {author} {\bibfnamefont
  {M.}~\bibnamefont {Kovacev}}, \bibinfo {author} {\bibfnamefont
  {U.}~\bibnamefont {Thumm}}, \bibinfo {author} {\bibfnamefont
  {I.}~\bibnamefont {Hartl}}, \bibinfo {author} {\bibfnamefont
  {R.}~\bibnamefont {Treusch}}, \bibinfo {author} {\bibfnamefont
  {R.}~\bibnamefont {Moshammer}}, \bibinfo {author} {\bibfnamefont
  {C.}~\bibnamefont {Ott}},\ and\ \bibinfo {author} {\bibfnamefont
  {T.}~\bibnamefont {Pfeifer}},\ }\bibfield  {title} {\bibinfo {title}
  {Time-resolving state-specific molecular dissociation with xuv broadband
  absorption spectroscopy},\ }\href
  {https://doi.org/doi:10.1126/sciadv.adk1482} {\bibfield  {journal} {\bibinfo
  {journal} {Science Advances}\ }\textbf {\bibinfo {volume} {9}},\ \bibinfo
  {pages} {eadk1482} (\bibinfo {year} {2023})}\BibitemShut {NoStop}%
\bibitem [{\citenamefont {Kleine}\ \emph {et~al.}(2022)\citenamefont {Kleine},
  \citenamefont {Winghart}, \citenamefont {Zhang}, \citenamefont {Richter},
  \citenamefont {Ekimova}, \citenamefont {Eckert}, \citenamefont {Vrakking},
  \citenamefont {Nibbering}, \citenamefont {Rouz\'ee},\ and\ \citenamefont
  {Grant}}]{kleine2022electronic}%
  \BibitemOpen
  \bibfield  {author} {\bibinfo {author} {\bibfnamefont {C.}~\bibnamefont
  {Kleine}}, \bibinfo {author} {\bibfnamefont {M.-O.}\ \bibnamefont
  {Winghart}}, \bibinfo {author} {\bibfnamefont {Z.-Y.}\ \bibnamefont {Zhang}},
  \bibinfo {author} {\bibfnamefont {M.}~\bibnamefont {Richter}}, \bibinfo
  {author} {\bibfnamefont {M.}~\bibnamefont {Ekimova}}, \bibinfo {author}
  {\bibfnamefont {S.}~\bibnamefont {Eckert}}, \bibinfo {author} {\bibfnamefont
  {M.~J.~J.}\ \bibnamefont {Vrakking}}, \bibinfo {author} {\bibfnamefont
  {E.~T.~J.}\ \bibnamefont {Nibbering}}, \bibinfo {author} {\bibfnamefont
  {A.}~\bibnamefont {Rouz\'ee}},\ and\ \bibinfo {author} {\bibfnamefont
  {E.~R.}\ \bibnamefont {Grant}},\ }\bibfield  {title} {\bibinfo {title}
  {Electronic state population dynamics upon ultrafast strong field ionization
  and fragmentation of molecular nitrogen},\ }\href
  {https://doi.org/10.1103/PhysRevLett.129.123002} {\bibfield  {journal}
  {\bibinfo  {journal} {Phys. Rev. Lett.}\ }\textbf {\bibinfo {volume} {129}},\
  \bibinfo {pages} {123002} (\bibinfo {year} {2022})}\BibitemShut {NoStop}%
\bibitem [{\citenamefont {Calegari}\ \emph {et~al.}(2016)\citenamefont
  {Calegari}, \citenamefont {Sansone}, \citenamefont {Stagira}, \citenamefont
  {Vozzi},\ and\ \citenamefont {Nisoli}}]{Calegari2016jpb}%
  \BibitemOpen
  \bibfield  {author} {\bibinfo {author} {\bibfnamefont {F.}~\bibnamefont
  {Calegari}}, \bibinfo {author} {\bibfnamefont {G.}~\bibnamefont {Sansone}},
  \bibinfo {author} {\bibfnamefont {S.}~\bibnamefont {Stagira}}, \bibinfo
  {author} {\bibfnamefont {C.}~\bibnamefont {Vozzi}},\ and\ \bibinfo {author}
  {\bibfnamefont {M.}~\bibnamefont {Nisoli}},\ }\bibfield  {title} {\bibinfo
  {title} {Advances in attosecond science},\ }\href
  {https://doi.org/10.1088/0953-4075/49/6/062001} {\bibfield  {journal}
  {\bibinfo  {journal} {Journal of Physics B: Atomic, Molecular and Optical
  Physics}\ }\textbf {\bibinfo {volume} {49}},\ \bibinfo {pages} {062001}
  (\bibinfo {year} {2016})}\BibitemShut {NoStop}%
\bibitem [{\citenamefont {Nisoli}(2019)}]{Nisoli19atto}%
  \BibitemOpen
  \bibfield  {author} {\bibinfo {author} {\bibfnamefont {M.}~\bibnamefont
  {Nisoli}},\ }\bibfield  {title} {\bibinfo {title} {The birth of
  attochemistry},\ }\href {https://doi.org/10.1364/OPN.30.7.000032} {\bibfield
  {journal} {\bibinfo  {journal} {Opt. Photon. News}\ }\textbf {\bibinfo
  {volume} {30}},\ \bibinfo {pages} {32} (\bibinfo {year} {2019})}\BibitemShut
  {NoStop}%
\bibitem [{\citenamefont {Kobayashi}\ and\ \citenamefont
  {Leone}(2022)}]{leone2022jcp}%
  \BibitemOpen
  \bibfield  {author} {\bibinfo {author} {\bibfnamefont {Y.}~\bibnamefont
  {Kobayashi}}\ and\ \bibinfo {author} {\bibfnamefont {S.~R.}\ \bibnamefont
  {Leone}},\ }\bibfield  {title} {\bibinfo {title} {{Characterizing coherences
  in chemical dynamics with attosecond time-resolved x-ray absorption
  spectroscopy}},\ }\href {https://doi.org/10.1063/5.0119942} {\bibfield
  {journal} {\bibinfo  {journal} {The Journal of Chemical Physics}\ }\textbf
  {\bibinfo {volume} {157}},\ \bibinfo {pages} {180901} (\bibinfo {year}
  {2022})}\BibitemShut {NoStop}%
\bibitem [{\citenamefont {Tong}\ \emph {et~al.}(2002)\citenamefont {Tong},
  \citenamefont {Zhao},\ and\ \citenamefont {Lin}}]{tong2002theory}%
  \BibitemOpen
  \bibfield  {author} {\bibinfo {author} {\bibfnamefont {X.~M.}\ \bibnamefont
  {Tong}}, \bibinfo {author} {\bibfnamefont {Z.~X.}\ \bibnamefont {Zhao}},\
  and\ \bibinfo {author} {\bibfnamefont {C.~D.}\ \bibnamefont {Lin}},\
  }\bibfield  {title} {\bibinfo {title} {Theory of molecular tunneling
  ionization},\ }\href {https://doi.org/10.1103/PhysRevA.66.033402} {\bibfield
  {journal} {\bibinfo  {journal} {Phys. Rev. A}\ }\textbf {\bibinfo {volume}
  {66}},\ \bibinfo {pages} {033402} (\bibinfo {year} {2002})}\BibitemShut
  {NoStop}%
\bibitem [{\citenamefont {Zhang}\ and\ \citenamefont
  {Zhao}(2015)}]{zhang2015slimp}%
  \BibitemOpen
  \bibfield  {author} {\bibinfo {author} {\bibfnamefont {B.}~\bibnamefont
  {Zhang}}\ and\ \bibinfo {author} {\bibfnamefont {Z.}~\bibnamefont {Zhao}},\
  }\bibfield  {title} {\bibinfo {title} {Slimp: Strong laser interaction model
  package for atoms and molecules},\ }\href
  {https://doi.org/https://doi.org/10.1016/j.cpc.2015.02.031} {\bibfield
  {journal} {\bibinfo  {journal} {Computer Physics Communications}\ }\textbf
  {\bibinfo {volume} {192}},\ \bibinfo {pages} {330} (\bibinfo {year}
  {2015})}\BibitemShut {NoStop}%
\bibitem [{\citenamefont {Wirth}\ \emph {et~al.}(2011)\citenamefont {Wirth},
  \citenamefont {Hassan}, \citenamefont {Grgura{\v s}}, \citenamefont {Gagnon},
  \citenamefont {Moulet}, \citenamefont {Luu}, \citenamefont {Pabst},
  \citenamefont {Santra}, \citenamefont {Alahmed}, \citenamefont {Azzeer},
  \citenamefont {Yakovlev}, \citenamefont {Pervak}, \citenamefont {Krausz},\
  and\ \citenamefont {Goulielmakis}}]{wirth2011synthesized}%
  \BibitemOpen
  \bibfield  {author} {\bibinfo {author} {\bibfnamefont {A.}~\bibnamefont
  {Wirth}}, \bibinfo {author} {\bibfnamefont {M.~T.}\ \bibnamefont {Hassan}},
  \bibinfo {author} {\bibfnamefont {I.}~\bibnamefont {Grgura{\v s}}}, \bibinfo
  {author} {\bibfnamefont {J.}~\bibnamefont {Gagnon}}, \bibinfo {author}
  {\bibfnamefont {A.}~\bibnamefont {Moulet}}, \bibinfo {author} {\bibfnamefont
  {T.~T.}\ \bibnamefont {Luu}}, \bibinfo {author} {\bibfnamefont
  {S.}~\bibnamefont {Pabst}}, \bibinfo {author} {\bibfnamefont
  {R.}~\bibnamefont {Santra}}, \bibinfo {author} {\bibfnamefont {Z.~A.}\
  \bibnamefont {Alahmed}}, \bibinfo {author} {\bibfnamefont {A.~M.}\
  \bibnamefont {Azzeer}}, \bibinfo {author} {\bibfnamefont {V.~S.}\
  \bibnamefont {Yakovlev}}, \bibinfo {author} {\bibfnamefont {V.}~\bibnamefont
  {Pervak}}, \bibinfo {author} {\bibfnamefont {F.}~\bibnamefont {Krausz}},\
  and\ \bibinfo {author} {\bibfnamefont {E.}~\bibnamefont {Goulielmakis}},\
  }\bibfield  {title} {\bibinfo {title} {Synthesized light transients},\ }\href
  {https://doi.org/10.1126/science.1210268} {\bibfield  {journal} {\bibinfo
  {journal} {Science}\ }\textbf {\bibinfo {volume} {334}},\ \bibinfo {pages}
  {195} (\bibinfo {year} {2011})}\BibitemShut {NoStop}%
\bibitem [{\citenamefont {Santra}\ \emph {et~al.}(2011)\citenamefont {Santra},
  \citenamefont {Yakovlev}, \citenamefont {Pfeifer},\ and\ \citenamefont
  {Loh}}]{santra2011pra}%
  \BibitemOpen
  \bibfield  {author} {\bibinfo {author} {\bibfnamefont {R.}~\bibnamefont
  {Santra}}, \bibinfo {author} {\bibfnamefont {V.~S.}\ \bibnamefont
  {Yakovlev}}, \bibinfo {author} {\bibfnamefont {T.}~\bibnamefont {Pfeifer}},\
  and\ \bibinfo {author} {\bibfnamefont {Z.-H.}\ \bibnamefont {Loh}},\
  }\bibfield  {title} {\bibinfo {title} {Theory of attosecond transient
  absorption spectroscopy of strong-field-generated ions},\ }\href
  {https://doi.org/10.1103/PhysRevA.83.033405} {\bibfield  {journal} {\bibinfo
  {journal} {Phys. Rev. A}\ }\textbf {\bibinfo {volume} {83}},\ \bibinfo
  {pages} {033405} (\bibinfo {year} {2011})}\BibitemShut {NoStop}%
\bibitem [{\citenamefont {Werme}\ \emph {et~al.}(1973)\citenamefont {Werme},
  \citenamefont {Grennberg}, \citenamefont {Nordgren}, \citenamefont
  {Nordling},\ and\ \citenamefont {Siegbahn}}]{werme1973fine}%
  \BibitemOpen
  \bibfield  {author} {\bibinfo {author} {\bibfnamefont {L.~O.}\ \bibnamefont
  {Werme}}, \bibinfo {author} {\bibfnamefont {B.}~\bibnamefont {Grennberg}},
  \bibinfo {author} {\bibfnamefont {J.}~\bibnamefont {Nordgren}}, \bibinfo
  {author} {\bibfnamefont {C.}~\bibnamefont {Nordling}},\ and\ \bibinfo
  {author} {\bibfnamefont {K.}~\bibnamefont {Siegbahn}},\ }\bibfield  {title}
  {\bibinfo {title} {Fine structure in the x-ray emission spectrum of {N}$_2$,
  compared with electron spectroscopy},\ }\href
  {https://doi.org/10.1038/242453a0} {\bibfield  {journal} {\bibinfo  {journal}
  {Nature}\ }\textbf {\bibinfo {volume} {242}},\ \bibinfo {pages} {453}
  (\bibinfo {year} {1973})}\BibitemShut {NoStop}%
\bibitem [{\citenamefont {Ehara}\ \emph {et~al.}(2006)\citenamefont {Ehara},
  \citenamefont {Nakatsuji}, \citenamefont {Matsumoto}, \citenamefont
  {Hatamoto}, \citenamefont {Liu}, \citenamefont {Lischke}, \citenamefont
  {Pr{\"u}mper}, \citenamefont {Tanaka}, \citenamefont {Makochekanwa},
  \citenamefont {Hoshino}, \citenamefont {Tanaka}, \citenamefont {Harries},
  \citenamefont {Tamenori},\ and\ \citenamefont {Ueda}}]{ehara2006symmetry}%
  \BibitemOpen
  \bibfield  {author} {\bibinfo {author} {\bibfnamefont {M.}~\bibnamefont
  {Ehara}}, \bibinfo {author} {\bibfnamefont {H.}~\bibnamefont {Nakatsuji}},
  \bibinfo {author} {\bibfnamefont {M.}~\bibnamefont {Matsumoto}}, \bibinfo
  {author} {\bibfnamefont {T.}~\bibnamefont {Hatamoto}}, \bibinfo {author}
  {\bibfnamefont {X.-J.}\ \bibnamefont {Liu}}, \bibinfo {author} {\bibfnamefont
  {T.}~\bibnamefont {Lischke}}, \bibinfo {author} {\bibfnamefont
  {G.}~\bibnamefont {Pr{\"u}mper}}, \bibinfo {author} {\bibfnamefont
  {T.}~\bibnamefont {Tanaka}}, \bibinfo {author} {\bibfnamefont
  {C.}~\bibnamefont {Makochekanwa}}, \bibinfo {author} {\bibfnamefont
  {M.}~\bibnamefont {Hoshino}}, \bibinfo {author} {\bibfnamefont
  {H.}~\bibnamefont {Tanaka}}, \bibinfo {author} {\bibfnamefont {J.~R.}\
  \bibnamefont {Harries}}, \bibinfo {author} {\bibfnamefont {Y.}~\bibnamefont
  {Tamenori}},\ and\ \bibinfo {author} {\bibfnamefont {K.}~\bibnamefont
  {Ueda}},\ }\bibfield  {title} {\bibinfo {title} {{Symmetry-dependent
  vibrational excitation in N 1s photoionization of N$_2$: Experiment and
  theory}},\ }\href {https://doi.org/10.1063/1.2181144} {\bibfield  {journal}
  {\bibinfo  {journal} {The Journal of Chemical Physics}\ }\textbf {\bibinfo
  {volume} {124}},\ \bibinfo {pages} {124311} (\bibinfo {year}
  {2006})}\BibitemShut {NoStop}%
\bibitem [{\citenamefont {Lindblad}\ \emph {et~al.}(2020)\citenamefont
  {Lindblad}, \citenamefont {Kjellsson}, \citenamefont {Couto}, \citenamefont
  {Timm}, \citenamefont {B\"ulow}, \citenamefont {Zamudio-Bayer}, \citenamefont
  {Lundberg}, \citenamefont {von Issendorff}, \citenamefont {Lau},
  \citenamefont {Sorensen}, \citenamefont {Carravetta}, \citenamefont
  {\AA{}gren},\ and\ \citenamefont {Rubensson}}]{lindblad2020x}%
  \BibitemOpen
  \bibfield  {author} {\bibinfo {author} {\bibfnamefont {R.}~\bibnamefont
  {Lindblad}}, \bibinfo {author} {\bibfnamefont {L.}~\bibnamefont {Kjellsson}},
  \bibinfo {author} {\bibfnamefont {R.~C.}\ \bibnamefont {Couto}}, \bibinfo
  {author} {\bibfnamefont {M.}~\bibnamefont {Timm}}, \bibinfo {author}
  {\bibfnamefont {C.}~\bibnamefont {B\"ulow}}, \bibinfo {author} {\bibfnamefont
  {V.}~\bibnamefont {Zamudio-Bayer}}, \bibinfo {author} {\bibfnamefont
  {M.}~\bibnamefont {Lundberg}}, \bibinfo {author} {\bibfnamefont
  {B.}~\bibnamefont {von Issendorff}}, \bibinfo {author} {\bibfnamefont
  {J.~T.}\ \bibnamefont {Lau}}, \bibinfo {author} {\bibfnamefont {S.~L.}\
  \bibnamefont {Sorensen}}, \bibinfo {author} {\bibfnamefont {V.}~\bibnamefont
  {Carravetta}}, \bibinfo {author} {\bibfnamefont {H.}~\bibnamefont
  {\AA{}gren}},\ and\ \bibinfo {author} {\bibfnamefont {J.-E.}\ \bibnamefont
  {Rubensson}},\ }\bibfield  {title} {\bibinfo {title} {X-ray absorption
  spectrum of the {N}$_{2}^{+}$ molecular ion},\ }\href
  {https://doi.org/10.1103/PhysRevLett.124.203001} {\bibfield  {journal}
  {\bibinfo  {journal} {Phys. Rev. Lett.}\ }\textbf {\bibinfo {volume} {124}},\
  \bibinfo {pages} {203001} (\bibinfo {year} {2020})}\BibitemShut {NoStop}%
\bibitem [{\citenamefont {Wei}\ \emph {et~al.}(2017)\citenamefont {Wei},
  \citenamefont {Li}, \citenamefont {Wang}, \citenamefont {See}, \citenamefont
  {Jhon}, \citenamefont {Zhang}, \citenamefont {Shi}, \citenamefont {Yang},\
  and\ \citenamefont {Loh}}]{loh2017nc}%
  \BibitemOpen
  \bibfield  {author} {\bibinfo {author} {\bibfnamefont {Z.}~\bibnamefont
  {Wei}}, \bibinfo {author} {\bibfnamefont {J.}~\bibnamefont {Li}}, \bibinfo
  {author} {\bibfnamefont {L.}~\bibnamefont {Wang}}, \bibinfo {author}
  {\bibfnamefont {S.~T.}\ \bibnamefont {See}}, \bibinfo {author} {\bibfnamefont
  {M.~H.}\ \bibnamefont {Jhon}}, \bibinfo {author} {\bibfnamefont
  {Y.}~\bibnamefont {Zhang}}, \bibinfo {author} {\bibfnamefont
  {F.}~\bibnamefont {Shi}}, \bibinfo {author} {\bibfnamefont {M.}~\bibnamefont
  {Yang}},\ and\ \bibinfo {author} {\bibfnamefont {Z.-H.}\ \bibnamefont
  {Loh}},\ }\bibfield  {title} {\bibinfo {title} {Elucidating the origins of
  multimode vibrational coherences of polyatomic molecules induced by intense
  laser fields},\ }\href {https://doi.org/10.1038/s41467-017-00848-2}
  {\bibfield  {journal} {\bibinfo  {journal} {Nature Communications}\ }\textbf
  {\bibinfo {volume} {8}},\ \bibinfo {pages} {735} (\bibinfo {year}
  {2017})}\BibitemShut {NoStop}%
\bibitem [{\citenamefont {Bakos}(1977)}]{bakos1977ac}%
  \BibitemOpen
  \bibfield  {author} {\bibinfo {author} {\bibfnamefont {J.}~\bibnamefont
  {Bakos}},\ }\bibfield  {title} {\bibinfo {title} {{AC} {S}tark effect and
  multiphoton processes in atoms},\ }\href
  {https://doi.org/https://doi.org/10.1016/0370-1573(77)90016-3} {\bibfield
  {journal} {\bibinfo  {journal} {Physics Reports}\ }\textbf {\bibinfo {volume}
  {31}},\ \bibinfo {pages} {209} (\bibinfo {year} {1977})}\BibitemShut
  {NoStop}%
\bibitem [{\citenamefont {Chini}\ \emph {et~al.}(2013)\citenamefont {Chini},
  \citenamefont {Wang}, \citenamefont {Cheng}, \citenamefont {Wu},
  \citenamefont {Zhao}, \citenamefont {Telnov}, \citenamefont {Chu},\ and\
  \citenamefont {Chang}}]{chini2013sr}%
  \BibitemOpen
  \bibfield  {author} {\bibinfo {author} {\bibfnamefont {M.}~\bibnamefont
  {Chini}}, \bibinfo {author} {\bibfnamefont {X.}~\bibnamefont {Wang}},
  \bibinfo {author} {\bibfnamefont {Y.}~\bibnamefont {Cheng}}, \bibinfo
  {author} {\bibfnamefont {Y.}~\bibnamefont {Wu}}, \bibinfo {author}
  {\bibfnamefont {D.}~\bibnamefont {Zhao}}, \bibinfo {author} {\bibfnamefont
  {D.~A.}\ \bibnamefont {Telnov}}, \bibinfo {author} {\bibfnamefont {S.-I.}\
  \bibnamefont {Chu}},\ and\ \bibinfo {author} {\bibfnamefont {Z.}~\bibnamefont
  {Chang}},\ }\bibfield  {title} {\bibinfo {title} {Sub-cycle oscillations in
  virtual states brought to light},\ }\href {https://doi.org/10.1038/srep01105}
  {\bibfield  {journal} {\bibinfo  {journal} {Scientific Reports}\ }\textbf
  {\bibinfo {volume} {3}},\ \bibinfo {pages} {1105} (\bibinfo {year}
  {2013})}\BibitemShut {NoStop}%
\bibitem [{\citenamefont {Zhou}\ \emph {et~al.}(2023)\citenamefont {Zhou},
  \citenamefont {Bao}, \citenamefont {Fan}, \citenamefont {Zhou}, \citenamefont
  {Gao}, \citenamefont {Zhong}, \citenamefont {Lin}, \citenamefont {Liu},
  \citenamefont {Yu}, \citenamefont {Tang}, \citenamefont {Meng}, \citenamefont
  {Duan},\ and\ \citenamefont {Zhou}}]{zhou23nature}%
  \BibitemOpen
  \bibfield  {author} {\bibinfo {author} {\bibfnamefont {S.}~\bibnamefont
  {Zhou}}, \bibinfo {author} {\bibfnamefont {C.}~\bibnamefont {Bao}}, \bibinfo
  {author} {\bibfnamefont {B.}~\bibnamefont {Fan}}, \bibinfo {author}
  {\bibfnamefont {H.}~\bibnamefont {Zhou}}, \bibinfo {author} {\bibfnamefont
  {Q.}~\bibnamefont {Gao}}, \bibinfo {author} {\bibfnamefont {H.}~\bibnamefont
  {Zhong}}, \bibinfo {author} {\bibfnamefont {T.}~\bibnamefont {Lin}}, \bibinfo
  {author} {\bibfnamefont {H.}~\bibnamefont {Liu}}, \bibinfo {author}
  {\bibfnamefont {P.}~\bibnamefont {Yu}}, \bibinfo {author} {\bibfnamefont
  {P.}~\bibnamefont {Tang}}, \bibinfo {author} {\bibfnamefont {S.}~\bibnamefont
  {Meng}}, \bibinfo {author} {\bibfnamefont {W.}~\bibnamefont {Duan}},\ and\
  \bibinfo {author} {\bibfnamefont {S.}~\bibnamefont {Zhou}},\ }\bibfield
  {title} {\bibinfo {title} {Pseudospin-selective floquet band engineering in
  black phosphorus},\ }\href {https://doi.org/10.1038/s41586-022-05610-3}
  {\bibfield  {journal} {\bibinfo  {journal} {Nature}\ }\textbf {\bibinfo
  {volume} {614}},\ \bibinfo {pages} {75} (\bibinfo {year} {2023})}\BibitemShut
  {NoStop}%
\bibitem [{\citenamefont {Lépine}\ \emph {et~al.}(2014)\citenamefont
  {Lépine}, \citenamefont {Ivanov},\ and\ \citenamefont
  {Vrakking}}]{vrakking2014np}%
  \BibitemOpen
  \bibfield  {author} {\bibinfo {author} {\bibfnamefont {F.}~\bibnamefont
  {Lépine}}, \bibinfo {author} {\bibfnamefont {M.~Y.}\ \bibnamefont
  {Ivanov}},\ and\ \bibinfo {author} {\bibfnamefont {M.~J.~J.}\ \bibnamefont
  {Vrakking}},\ }\bibfield  {title} {\bibinfo {title} {Attosecond molecular
  dynamics: fact or fiction?},\ }\href
  {https://doi.org/10.1038/nphoton.2014.25} {\bibfield  {journal} {\bibinfo
  {journal} {Nature Photonics}\ }\textbf {\bibinfo {volume} {8}},\ \bibinfo
  {pages} {195} (\bibinfo {year} {2014})}\BibitemShut {NoStop}%
\bibitem [{\citenamefont {Driver}\ \emph {et~al.}(2024)\citenamefont {Driver},
  \citenamefont {Mountney}, \citenamefont {Wang}, \citenamefont {Ortmann},
  \citenamefont {Al-Haddad}, \citenamefont {Berrah}, \citenamefont {Bostedt},
  \citenamefont {Champenois}, \citenamefont {DiMauro}, \citenamefont {Duris},
  \citenamefont {Garratt}, \citenamefont {Glownia}, \citenamefont {Guo},
  \citenamefont {Haxton}, \citenamefont {Isele}, \citenamefont {Ivanov},
  \citenamefont {Ji}, \citenamefont {Kamalov}, \citenamefont {Li},
  \citenamefont {Lin}, \citenamefont {Marangos}, \citenamefont {Obaid},
  \citenamefont {O’Neal}, \citenamefont {Rosenberger}, \citenamefont
  {Shivaram}, \citenamefont {Wang}, \citenamefont {Walter}, \citenamefont
  {Wolf}, \citenamefont {Wörner}, \citenamefont {Zhang}, \citenamefont
  {Bucksbaum}, \citenamefont {Kling}, \citenamefont {Landsman}, \citenamefont
  {Lucchese}, \citenamefont {Emmanouilidou}, \citenamefont {Marinelli},\ and\
  \citenamefont {Cryan}}]{james2024nature}%
  \BibitemOpen
  \bibfield  {author} {\bibinfo {author} {\bibfnamefont {T.}~\bibnamefont
  {Driver}}, \bibinfo {author} {\bibfnamefont {M.}~\bibnamefont {Mountney}},
  \bibinfo {author} {\bibfnamefont {J.}~\bibnamefont {Wang}}, \bibinfo {author}
  {\bibfnamefont {L.}~\bibnamefont {Ortmann}}, \bibinfo {author} {\bibfnamefont
  {A.}~\bibnamefont {Al-Haddad}}, \bibinfo {author} {\bibfnamefont
  {N.}~\bibnamefont {Berrah}}, \bibinfo {author} {\bibfnamefont
  {C.}~\bibnamefont {Bostedt}}, \bibinfo {author} {\bibfnamefont {E.~G.}\
  \bibnamefont {Champenois}}, \bibinfo {author} {\bibfnamefont {L.~F.}\
  \bibnamefont {DiMauro}}, \bibinfo {author} {\bibfnamefont {J.}~\bibnamefont
  {Duris}}, \bibinfo {author} {\bibfnamefont {D.}~\bibnamefont {Garratt}},
  \bibinfo {author} {\bibfnamefont {J.~M.}\ \bibnamefont {Glownia}}, \bibinfo
  {author} {\bibfnamefont {Z.}~\bibnamefont {Guo}}, \bibinfo {author}
  {\bibfnamefont {D.}~\bibnamefont {Haxton}}, \bibinfo {author} {\bibfnamefont
  {E.}~\bibnamefont {Isele}}, \bibinfo {author} {\bibfnamefont
  {I.}~\bibnamefont {Ivanov}}, \bibinfo {author} {\bibfnamefont
  {J.}~\bibnamefont {Ji}}, \bibinfo {author} {\bibfnamefont {A.}~\bibnamefont
  {Kamalov}}, \bibinfo {author} {\bibfnamefont {S.}~\bibnamefont {Li}},
  \bibinfo {author} {\bibfnamefont {M.-F.}\ \bibnamefont {Lin}}, \bibinfo
  {author} {\bibfnamefont {J.~P.}\ \bibnamefont {Marangos}}, \bibinfo {author}
  {\bibfnamefont {R.}~\bibnamefont {Obaid}}, \bibinfo {author} {\bibfnamefont
  {J.~T.}\ \bibnamefont {O’Neal}}, \bibinfo {author} {\bibfnamefont
  {P.}~\bibnamefont {Rosenberger}}, \bibinfo {author} {\bibfnamefont {N.~H.}\
  \bibnamefont {Shivaram}}, \bibinfo {author} {\bibfnamefont {A.~L.}\
  \bibnamefont {Wang}}, \bibinfo {author} {\bibfnamefont {P.}~\bibnamefont
  {Walter}}, \bibinfo {author} {\bibfnamefont {T.~J.~A.}\ \bibnamefont {Wolf}},
  \bibinfo {author} {\bibfnamefont {H.~J.}\ \bibnamefont {Wörner}}, \bibinfo
  {author} {\bibfnamefont {Z.}~\bibnamefont {Zhang}}, \bibinfo {author}
  {\bibfnamefont {P.~H.}\ \bibnamefont {Bucksbaum}}, \bibinfo {author}
  {\bibfnamefont {M.~F.}\ \bibnamefont {Kling}}, \bibinfo {author}
  {\bibfnamefont {A.~S.}\ \bibnamefont {Landsman}}, \bibinfo {author}
  {\bibfnamefont {R.~R.}\ \bibnamefont {Lucchese}}, \bibinfo {author}
  {\bibfnamefont {A.}~\bibnamefont {Emmanouilidou}}, \bibinfo {author}
  {\bibfnamefont {A.}~\bibnamefont {Marinelli}},\ and\ \bibinfo {author}
  {\bibfnamefont {J.~P.}\ \bibnamefont {Cryan}},\ }\bibfield  {title} {\bibinfo
  {title} {Attosecond delays in x-ray molecular ionization},\ }\href
  {https://doi.org/10.1038/s41586-024-07771-9} {\bibfield  {journal} {\bibinfo
  {journal} {Nature}\ }\textbf {\bibinfo {volume} {632}},\ \bibinfo {pages}
  {762} (\bibinfo {year} {2024})}\BibitemShut {NoStop}%
\bibitem [{\citenamefont {Marroux}\ \emph {et~al.}(2020)\citenamefont
  {Marroux}, \citenamefont {Fidler}, \citenamefont {Ghosh}, \citenamefont
  {Kobayashi}, \citenamefont {Gokhberg}, \citenamefont {Kuleff}, \citenamefont
  {Leone},\ and\ \citenamefont {Neumark}}]{neumark2020nc}%
  \BibitemOpen
  \bibfield  {author} {\bibinfo {author} {\bibfnamefont {H.~J.~B.}\
  \bibnamefont {Marroux}}, \bibinfo {author} {\bibfnamefont {A.~P.}\
  \bibnamefont {Fidler}}, \bibinfo {author} {\bibfnamefont {A.}~\bibnamefont
  {Ghosh}}, \bibinfo {author} {\bibfnamefont {Y.}~\bibnamefont {Kobayashi}},
  \bibinfo {author} {\bibfnamefont {K.}~\bibnamefont {Gokhberg}}, \bibinfo
  {author} {\bibfnamefont {A.~I.}\ \bibnamefont {Kuleff}}, \bibinfo {author}
  {\bibfnamefont {S.~R.}\ \bibnamefont {Leone}},\ and\ \bibinfo {author}
  {\bibfnamefont {D.~M.}\ \bibnamefont {Neumark}},\ }\bibfield  {title}
  {\bibinfo {title} {Attosecond spectroscopy reveals alignment dependent
  core-hole dynamics in the icl molecule},\ }\href
  {https://doi.org/10.1038/s41467-020-19496-0} {\bibfield  {journal} {\bibinfo
  {journal} {Nature Communications}\ }\textbf {\bibinfo {volume} {11}},\
  \bibinfo {pages} {5810} (\bibinfo {year} {2020})}\BibitemShut {NoStop}%
\end{thebibliography}%
%apsrev4-2.bst 2019-01-14 (MD) hand-edited version of apsrev4-1.bst
%Control: key (0)
%Control: author (8) initials jnrlst
%Control: editor formatted (1) identically to author
%Control: production of article title (0) allowed
%Control: page (0) single
%Control: year (1) truncated
%Control: production of eprint (0) enabled
\providecommand{\noopsort}[1]{}\providecommand{\singleletter}[1]{#1}%

\end{document}